\input harvmac
\overfullrule=0pt
\parindent 25pt
\tolerance=10000

\input epsf

\newcount\figno
\figno=0
\def\fig#1#2#3{
\par\begingroup\parindent=0pt\leftskip=1cm\rightskip=1cm\parindent=0pt
\baselineskip=11pt
\global\advance\figno by 1
\midinsert
\epsfxsize=#3
\centerline{\epsfbox{#2}}
\vskip 12pt
{\bf Fig.\ \the\figno: } #1\par
\endinsert\endgroup\par
}
\def\figlabel#1{\xdef#1{\the\figno}}
\def\encadremath#1{\vbox{\hrule\hbox{\vrule\kern8pt\vbox{\kern8pt
\hbox{$\displaystyle #1$}\kern8pt}
\kern8pt\vrule}\hrule}}

\def\half{{\textstyle{1\over2}}}

\def\apm{{\alpha^{\prime}}}

\def\half{{1\over 2}}
 
 \def\m{{\mu}}

 \def\ep{{\epsilon}}

 \def\t{{\theta}}
 \def\a{{\alpha}}
 
 \def\frac#1#2{{#1\over #2}}

 \def\s{{\sigma}}

 \def\CO{{\cal O}}
 
 \def\Ph{{\Phi }}

 \def\p{\partial}

 \def\apm{{\alpha'}}
 \def\r{\rightarrow}

 \def\al{\alpha'}
 \def\de{\partial}
 
 \def\we{\wedge}
 \def\lr{\leftrightarrow}
 \def\f {\frac}
 \def\ti{\tilde}
 \def\ap{\alpha}

 \def\ddd{\cdot\cdot\cdot}
 
 \def\la{\langle}
 \def\lb{\rangle}
 \def\ep{\epsilon}

 \def\ov{\overline}
 \def\sq{\sqrt}

\lref\AdPoSi{
A.~Adams, J.~Polchinski and E.~Silverstein,
``Don't panic! Closed string tachyons in ALE space-times,''
JHEP {\bf 0110} (2001) 029
[arXiv:hep-th/0108075].}

\lref\DabholkarIF{
A.~Dabholkar,
``Tachyon condensation and black hole entropy,''
arXiv:hep-th/0111004.
}

\lref\chicago{ J.~A.~Harvey, D.~Kutasov, E.~J.~Martinec, G.~Moore,
``Localized tachyons and RG flows'', arXiv:hep-th/0111154.
}

\lref\bega{
O.~Bergman and M.~R.~Gaberdiel,
``Dualities of type 0 strings,''
JHEP {\bf 9907} (1999) 022
[arXiv:hep-th/9906055].}

\lref\Va{
C.~Vafa,
``Mirror symmetry and closed string tachyon condensation,''
arXiv:hep-th/0111051;
A.~Dabholkar and C.~Vafa,
``tt* geometry and closed string tachyon potential,''
JHEP {\bf 0202} (2002) 008
[arXiv:hep-th/0111155].}

\lref\DaGuHeMi{
J.~R.~David, M.~Gutperle, M.~Headrick and S.~Minwalla,
``Closed string tachyon condensation on twisted circles,''
JHEP {\bf 0202}, 041 (2002)
[arXiv:hep-th/0111212].
}

\lref\GuHeMiSc{
M.~Gutperle, M.~Headrick, S.~Minwalla and V.~Schomerus,
``Space-time energy decreases under world-sheet RG flow,''
arXiv:hep-th/0211063.}

\lref\HoVa{
K.~Hori and C.~Vafa,
``Mirror symmetry,''
arXiv:hep-th/0002222.}

\lref\HoIqVa{
K.~Hori, A.~Iqbal and C.~Vafa,
``D-branes and mirror symmetry,''
arXiv:hep-th/0005247.}

\lref\HoKa{
K.~Hori and A.~Kapustin,
``Duality of the fermionic 2d black hole and N = 2 Liouville 
theory as  mirror symmetry,''
JHEP {\bf 0108} (2001) 045
[arXiv:hep-th/0104202].}

\lref\WittenBSFT{
E.~Witten,
``On background independent open string field theory,''
Phys.\ Rev.\ D {\bf 46} (1992) 5467
[arXiv:hep-th/9208027];

E.~Witten,
``Some computations in background independent off-shell string theory,''
Phys.\ Rev.\ D {\bf 47} (1993) 3405
[arXiv:hep-th/9210065].}

\lref\BSFT{
A.~A.~Gerasimov and S.~L.~Shatashvili,
``On exact tachyon potential in open string field theory,''
JHEP {\bf 0010} (2000) 034
[arXiv:hep-th/0009103];

D.~Kutasov, M.~Marino and G.~W.~Moore,
``Some exact results on tachyon condensation in string field theory,''
JHEP {\bf 0010} (2000) 045
[arXiv:hep-th/0009148];

D.~Kutasov, M.~Marino and G.~W.~Moore,
``Remarks on tachyon condensation in superstring field theory,''
arXiv:hep-th/0010108.}

\lref\DD{
P.~Kraus and F.~Larsen,
``Boundary string field theory of the DD-bar system,''
Phys.\ Rev.\ D {\bf 63} (2001) 106004
[arXiv:hep-th/0012198];

T.~Takayanagi, S.~Terashima and T.~Uesugi,
``Brane-antibrane action from boundary string field theory,''
JHEP {\bf 0103} (2001) 019
[arXiv:hep-th/0012210].}

\lref\Ho{
K.~Hori,
``Linear models of supersymmetric D-branes,''
arXiv:hep-th/0012179.}

\lref\ta{
T.~Takayanagi,
``Holomorphic tachyons and fractional D-branes,''
Nucl.\ Phys.\ B {\bf 603} (2001) 259
[arXiv:hep-th/0103021];
T.~Takayanagi,
``Tachyon condensation on orbifolds and McKay correspondence,''
Phys.\ Lett.\ B {\bf 519} (2001) 137
[arXiv:hep-th/0106142].}

\lref\DoMo{
M.~R.~Douglas and G.~W.~Moore,
``D-branes, Quivers, and ALE Instantons,''
arXiv:hep-th/9603167.}

\lref\DiDoGo{
D.~E.~Diaconescu, M.~R.~Douglas and J.~Gomis,
``Fractional branes and wrapped branes,''
JHEP {\bf 9802} (1998) 013
[arXiv:hep-th/9712230].}

\lref\DiGo{
D.~E.~Diaconescu and J.~Gomis,
``Fractional branes and boundary states in orbifold theories,''
JHEP {\bf 0010} (2000) 001
[arXiv:hep-th/9906242].}

\lref\TaUeo{
T.~Takayanagi and T.~Uesugi,
``Orbifolds as Melvin geometry,''
JHEP {\bf 0112} (2001) 004
[arXiv:hep-th/0110099].}

\lref\TaUet{
T.~Takayanagi and T.~Uesugi,
``D-branes in Melvin background,''
JHEP {\bf 0111} (2001) 036
[arXiv:hep-th/0110200];
T.~Takayanagi and T.~Uesugi,
``Flux stabilization of D-branes in NSNS Melvin background,''
Phys.\ Lett.\ B {\bf 528} (2002) 156
[arXiv:hep-th/0112199].}

\lref\DuMo{
E.~Dudas and J.~Mourad,
``D-branes in string theory Melvin backgrounds,''
Nucl.\ Phys.\ B {\bf 622} (2002) 46
[arXiv:hep-th/0110186].}

\lref\CoGu{
M.~S.~Costa and M.~Gutperle,
``The Kaluza-Klein Melvin solution in M-theory,''
JHEP {\bf 0103} (2001) 027
[arXiv:hep-th/0012072].}

\lref\GuSt{
M.~Gutperle and A.~Strominger,
``Fluxbranes in string theory,''
JHEP {\bf 0106} (2001) 035
[arXiv:hep-th/0104136].}

\lref\Gu{
M.~Gutperle,
``A note on perturbative and nonperturbative instabilities of 
twisted  circles,''
Phys.\ Lett.\ B {\bf 545}, 379 (2002)
[arXiv:hep-th/0207131].
}

\lref\Da{
J.~R.~David,
``Unstable magnetic fluxes in heterotic string theory,''
JHEP {\bf 0209}, 006 (2002)
[arXiv:hep-th/0208011].
}

\lref\HeKaLaMc{
S.~Hellerman, S.~Kachru, A.~E.~Lawrence and J.~McGreevy,
``Linear sigma models for open strings,''
JHEP {\bf 0207} (2002) 002
[arXiv:hep-th/0109069].}

\lref\EgSu{
T.~Eguchi and Y.~Sugawara,
``D-branes in singular Calabi-Yau n-fold and N = 2 Liouville theory,''
Nucl.\ Phys.\ B {\bf 598} (2001) 467
[arXiv:hep-th/0011148].}

\lref\OoOzYi{
H.~Ooguri, Y.~Oz and Z.~Yin,
``D-branes on Calabi-Yau spaces and their mirrors,''
Nucl.\ Phys.\ B {\bf 477} (1996) 407
[arXiv:hep-th/9606112].}

\lref\RussoTF{
J.~G.~Russo and A.~A.~Tseytlin,
 ``Magnetic backgrounds and tachyonic instabilities in closed superstring  
theory and M-theory,''
Nucl.\ Phys.\ B {\bf 611}, 93 (2001)
[arXiv:hep-th/0104238].
}

\lref\WittenYC{
E.~Witten,
``Phases of N = 2 theories in two dimensions,''
Nucl.\ Phys.\ B {\bf 403}, 159 (1993)
[arXiv:hep-th/9301042].
}

\lref\SuyamaGD{
T.~Suyama,
``Properties of string theory on Kaluza-Klein Melvin background,''
JHEP {\bf 0207}, 015 (2002)
[arXiv:hep-th/0110077].
}

\lref\TseytlinZV{
A.~A.~Tseytlin,
``Closed superstrings in magnetic flux tube background,''
Nucl.\ Phys.\ Proc.\ Suppl.\  {\bf 49}, 338 (1996)
[arXiv:hep-th/9510041].
}

\lref\TseytlinEI{
A.~A.~Tseytlin,
``Melvin solution in string theory,''
Phys.\ Lett.\ B {\bf 346}, 55 (1995)
[arXiv:hep-th/9411198].
}

\lref\wittenbh{
E.~Witten,
``On string theory and black holes,''
Phys.\ Rev.\ D {\bf 44}, 314 (1991).
}

\lref\tse{
J.~G.~Russo and A.~A.~Tseytlin,
``Magnetic flux tube models in superstring theory,''
Nucl.\ Phys.\ B {\bf 461}, 131 (1996)
[arXiv:hep-th/9508068].
}

\lref\piljin{
Y.~Michishita and P.~Yi,
``D-brane probe and closed string tachyons,''
Phys.\ Rev.\ D {\bf 65}, 086006 (2002)
[arXiv:hep-th/0111199].
}

\lref\MaMo{
E.~J.~Martinec and G.~Moore,
``On decay of K-theory,''
arXiv:hep-th/0212059.
}

\lref\Goa{
S.~Govindarajan and T.~Jayaraman,
``On the Landau-Ginzburg description of boundary CFTs and special 
Lagrangian submanifolds,''
JHEP {\bf 0007}, 016 (2000)
[arXiv:hep-th/0003242].
}

\lref\Gob{
S.~Govindarajan, T.~Jayaraman and T.~Sarkar,
``On D-branes from gauged linear sigma models,''
Nucl.\ Phys.\ B {\bf 593}, 155 (2001)
[arXiv:hep-th/0007075].
}

\lref\Goc{
S.~Govindarajan and T.~Jayaraman,
``Boundary fermions, coherent sheaves and D-branes on Calabi-Yau  
manifolds,''
Nucl.\ Phys.\ B {\bf 618}, 50 (2001)
[arXiv:hep-th/0104126].
}

\lref\HeMc{
S.~Hellerman and J.~McGreevy,
``Linear sigma model toolshed for D-brane physics,''
JHEP {¥bf 0110}, 002 (2001)
[arXiv:hep-th/0104100].
}

\lref\SuYa{
K.~Sugiyama and S.~Yamaguchi,
``D-branes on a noncompact singular Calabi-Yau manifold,''
JHEP {\bf 0102}, 015 (2001)
[arXiv:hep-th/0011091].
}

\lref\DoFi{
M.~R.~Douglas and B.~Fiol,
``D-branes and discrete torsion. II,''
arXiv:hep-th/9903031.
}

\lref\Vaq{
C.~Vafa,
``Quantum Symmetries Of String Vacua,''
Mod.\ Phys.\ Lett.\ A {\bf 4}, 1615 (1989).
}

\lref\Bi{
M.~Billo, B.~Craps and F.~Roose,
``Orbifold boundary states from Cardy's condition,''
JHEP {\bf 0101}, 038 (2001)
[arXiv:hep-th/0011060].
}

\lref\CallanBC{
C.~G.~Callan, C.~Lovelace, C.~R.~Nappi and S.~A.~Yost,
``String Loop Corrections To Beta Functions,''
Nucl.\ Phys.\ B {\bf 288}, 525 (1987).
}

\lref\SenSM{
A.~Sen,
``Tachyon condensation on the brane antibrane system,''
JHEP {\bf 9808}, 012 (1998)
[arXiv:hep-th/9805170].
}

\lref\Martinec{
E.~J.~Martinec,
``Defects, decay, and dissipated states,''
arXiv:hep-th/0210231.
}

\lref\YHE{
Y.~H.~He,
``Closed string tachyons, non-supersymmetric orbifolds and 
generalised  McKay correspondence,''
arXiv:hep-th/0301162.
}

\baselineskip 18pt plus 2pt minus 2pt

\Title{\vbox{\baselineskip12pt \hbox{hep-th/0307248}\hbox{HUTP-03/A046}
  }}
{\vbox{\centerline{Evolution of D-branes} 
\centerline{Under Closed String Tachyon 
Condensation}}}
\centerline{Shiraz Minwalla and Tadashi Takayanagi }

\medskip\centerline{ Jefferson Physical Laboratory}
\centerline{Harvard University}
\centerline{Cambridge, MA 02138}

\vskip .1in \centerline{\bf Abstract}

We study the evolution of stable D-branes of ${\bf C/Z}_n$ and 
twisted circle
theories in the process of closed string tachyon condensation. 
We interpret 
the fractional branes in these backgrounds as type II branes wrapping 
(`blown up') cycles, and trace the evolution of
the corresponding cycles under tachyon condensation. We also
study RG flows of the corresponding ${\cal N}=2$ boundary conformal field
theories. We find flows along which fractional D-branes either 
disappear or evolve into other fractional D-branes, and other flows along
which bulk branes either disappear or evolve into stable branes.

\noblackbox

\Date{July 2003}

\listtoc
\writetoc

\newsec{Introduction}

Consider a classical scalar field theory with action 
\eqn\actwc{S= \int d^dx {1\over 2 g^2} \left[(\p_\mu \ph)^2 +V(\ph)\right]}
where $V(\ph)$ takes the form shown in Fig. 1. A glance at this action 
carries a lot of information. For instance, it is immediately clear that
this system has at least two classically stable static 
backgrounds ($\ph= a$ and
$\ph=c$) and two unstable backgrounds ($\ph=0$ and $\ph=b$). The
spectrum around the unstable solution $\ph=0$ includes a tachyon; the end
point of tachyon condensation process is a gas of radiation about 
the stable vacuum $\ph=a$. 
\fig{Configuration space of scalar field theory with potential.}
{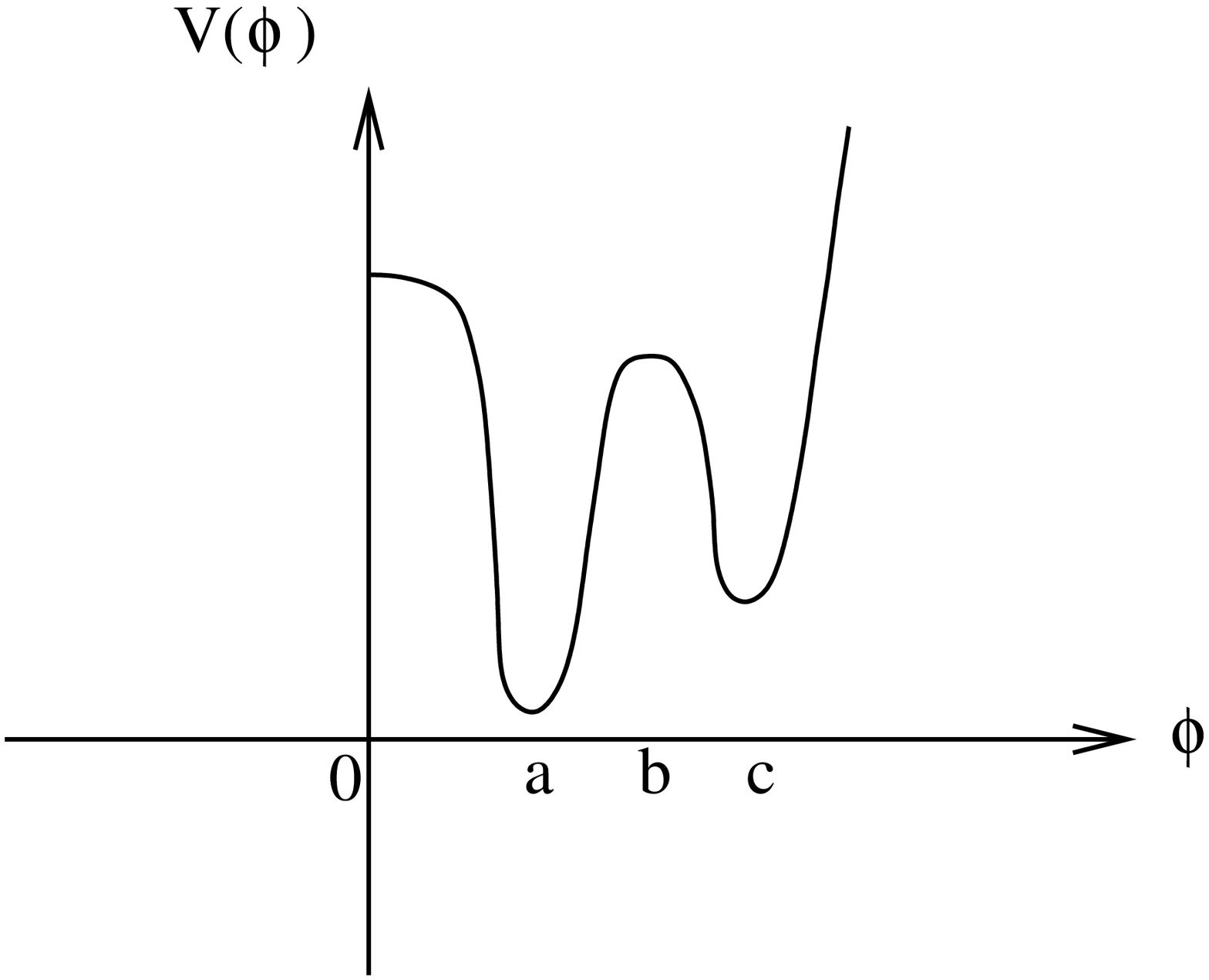}{3.0truein}

It is instructive to contrast our understanding of the model \actwc\ with
our current understanding of  string theory. We know a great deal 
about the properties of several particular classical solutions of string 
theory but lack intuition on the global structure of the configuration
space. It is not even clear whether all solutions (e.g. the Bosonic String)
are part of the same configuration space.

The recent study of closed string tachyon condensation has begun to address 
this issue. For example, Adams, Polchinski and Silverstein (APS) \AdPoSi\ 
have argued that unstable 
${\bf C/ Z}_n$ 
orbifolds of type II theory decay to type II theory in flat space, a
process roughly analogous to the decay of $\ph=0$ to 
$\ph=a$ in the model \actwc\ described above (see also \DabholkarIF 
\Va \chicago \Martinec). 
Similarly, generalizations of this study indicate that twisted circles or 
Melvin compactifications of type II theory \tse\ 
decay to supersymmetric 
circle compactifications of flat space \DaGuHeMi\ 
(for recent discussions of closed string tachyon condensation
in the twisted circle see also e.g. \GuSt \RussoTF \SuyamaGD \Gu \Da). 
These decay processes access otherwise unexplored regions of the 
configuration space of string theory, and deserve to be studied in greater 
detail. D-branes have often proved to be useful probes in string theory and
the unstable backgrounds we study host a rich spectrum of stable D-branes.
In this paper we will study how these stable branes evolve in the process
of closed string tachyon condensation.\foot{See \MaMo\ for a discussion 
of similar issues in the context of nonsupersymmetric 
${\bf C}^2 /{\bf Z}_n$.
orbifolds.}

Our study will focus largely on twisted circle theories. These rather 
versatile models appear in a two parameter family labeled by the 
size of the 
twisted circle $R$ and the angular twist $2\pi (1-{1\over n})$ in 
an auxiliary 2-plane,  upon traversing 
this circle . As we review in 
section 2 below, twisted circles reduce to ${\bf C/Z}_n$ orbifolds 
in the limit\foot{Intriguingly 
they also reduce to orbifolds of type 0 theory
in the limit $n \to \infty$; however the smoothness of this limit is open
to debate.} $R \to 0$ \TaUeo, and so may be thought of as a
desingularization or `blow up' of ${\bf C / Z}_n$ (in particular 
the ${\bf C/Z}_n$ fractional 0-brane may be `blown up' into a wrapped 
twisted
circle 1-brane). Consequently, twisted circle dynamics reduces to 
 ${\bf C /Z}_n$ dynamics upon taking the appropriate limit. 

We study the evolution of D-branes under tachyon condensation using two 
complementary approaches. Our first approach uses the remarkable fact that 
the Euclidean (more precisely Liouville) evolution of twisted 
circle models is approximately captured by an easily obtained and completely 
smooth interpolating geometry, a space that is similar in some respects 
to Taub Nut space. In order to take advantage of this fact, we
interpret the D-branes on twisted circle backgrounds as usual type II 
branes wrapping various geometric cycles of the twisted circle background. 
We then follow the evolution of these cycles to late Euclidean `time' in the
interpolating RG-flow geometry. We find that the cycles wrapped by twisted 
circle D-branes always evolve, under RG flow in one of three distinct ways.
\item{1.} In some cases the relevant cycle remains nonsingular through the 
process of RG flow and smoothly evolves into a nontrivial cycle in the IR 
geometry. This allows us to unambiguously
trace the evolution of the corresponding D-brane under tachyon condensation.
For example we conclude that  the twisted circle 3-brane evolves into the 
usual type II 3-brane (a similar result is true of the 2-brane of 
${\bf C/ Z}_n$).
\item{2.} In other cases, the relevant cycle shrinks to zero size at a 
particular Euclidean time, and remains trivial through the  subsequent 
RG flow. This happens only at the 
fixed point (or the origin) of the orbifold.
In these cases (which 
includes the twisted circle fractional 1-brane and so the ${\bf C/ Z}_n$ 
fractional 0-brane) we conjecture the existence of flows along which the 
corresponding D-brane simply disappears at some finite time during RG flow. 
\item{3.} Most interestingly, in some cases the relevant cycle degenerates
at exactly one point in the full RG flow geometry (intuitively, this 
cycle is non vanishing both before and after the singular point $r=0$).
In such cases we conjecture the existence of two qualitatively distinct 
RG-flows (precisely which one is implemented depends on the details of the 
initial conditions). Along the first kind of flow the corresponding D-brane
disappears when its cycle degenerates. Along the second kind of flow
the D-brane evolves, through the singularity, and eventually wraps the 
corresponding smooth cycle in the supersymmetric space. For example, we 
conjecture that the twisted circle bulk 1-brane (resp the ${\bf C/Z}_n$ 
bulk 0-brane) can either disappear near the origin, 
or evolve into a 1-brane wrapping the 
supersymmetric circle (resp bulk 0-brane on ${\bf C}$) under the process of 
tachyon condensation. 

{~~~}

Geometrical considerations also lead us to conjecture 
the existence of flows in which certain bulk branes (e.g. a D1-brane
wrapped on a compactified circle in flat space) are created out of nothing, 
in the process of tachyon 
condensation. Finally the flows we have 
described above may be superposed to produce more complicated flows (e.g. 
one in which a fractional brane is disappears, while at the same time 
a bulk brane is created out of nothing).

We are able to confirm some of these geometrically inspired 
conjectures using a more rigorous approach. Recall that 
\Va\ and \DaGuHeMi\ were able to obtain exact results on closed string
tachyon condensation 
by studying ${\cal N}=2$ supersymmetric renormalization group flows in 
Gauged Linear 
Sigma Models (GLSMs). Introducing D-branes into the story 
corresponds to putting the GLSMs on the disk with appropriate boundary 
conditions and adding relevant boundary degrees of freedom 
(`Chan Paton factors') \HoIqVa \Gob \Ho. When the boundary 
conditions and interactions in question also preserve ${\cal N}=2$ 
supersymmetry, it is not difficult to construct RG flows that follow the 
evolution of these boundary conditions to late Euclidean times. 
Using this method, we are able 
to construct examples of all the flows described in items 1-3 above, 
including both kinds of flows described in item 3. As one check on the formal 
reasoning in this part of the paper, we present (following 
\AdPoSi\ and \GuHeMiSc\ in the absence of boundaries) an explicit solution 
to the 1-loop beta function equations that describes the ${\bf C/Z}_n$ 2-brane 
relaxing to the supersymmetric 2-brane in flat space.

As we have noted, D-branes disappear along some of the flows 
described above. Interestingly this disappearance occurs, roughly, via a 
process of  open-string tachyon condensation (even though the branes under 
study were stable in the UV). This open string instability is  triggered 
(at some finite Euclidean time) by the closed string tachyon condensation 
process (see sections 5 and 7 for more details).  

We have, so far, described the evolution of D-branes under generic bulk 
RG flows, along which twisted circles and ${\bf C/ Z}_n$ decay to flat space. 
As is well known, world sheet $N=2$ supersymmetry guarantees the 
existence of 
more fine tuned RG flows in both these models; for instance flows from  
${\bf C}/{\bf Z}_n$ to ${\bf C}/{\bf Z}_m\ \ (n>m)$. It is not difficult 
to repeat the analysis described above on these fine tuned flows. As above 
we find the geometrically motivated conjectures as to the fate of these 
D-branes are always borne out by a more rigorous analysis of these 
flows. In 
particular, the flows describing the evolution of a single fractional 
D-branes 
are always of type described in item 2 above. Some multi-fractional branes
fall into the category of item 3; such branes either disappear or 
evolve into specific other fractional branes under closed string tachyon 
condensation. The story with bulk branes is unchanged. The analysis 
of these special flows is particularly useful as it permits a better 
understand of how geometrical information is encoded in corresponding 
boundary interactions due to open string tachyon fields.

This paper is organized as follows. In section 2 we review the closed
string backgrounds of interest to this paper, namely  twisted circles, 
${\bf C}/{\bf Z}_n$  orbifolds and type 0 theory. In section 3 we review 
the tachyon condensation of twisted circle models \DaGuHeMi\ and  
derive a smooth interpolating Euclidean 4-geometry that approximately
captures this tachyon condensation, and study it in detail. 
In section 4 we review the D-branes in twisted circle backgrounds, and 
interpret them as usual D-branes wrapping nontrivial cycles in the twisted
circle background. We also demonstrate that these branes reduce, in the 
appropriate limits, to the D-branes of ${\bf C/Z}_n$ and type 0 theory. 
In section 5 we study the evolution of stable D-branes under tachyon 
condensation in the models introduced above from two different viewpoints. 
In section 6 we consider evolution of D-branes under closed
tachyon condensation via more general 
decay channels of the twisted circles and orbifolds.
In Appendix A we review the so called `interpolating 
orbifold' (which interpolates  between type II and type 0 string theory).
In Appendix B we give a detailed review of closed string tachyon 
condensation in ${\bf C/Z}_n$ and its description by gauged 
linear sigma model with some new results.
In Appendix C we compute the classical sigma model metric of gauged 
linear sigma model for the twisted circle. In Appendix D we
discuss D-branes in the Landau-Ginzburg (LG) models 
which are dual to our models.
In Appendix E we present 
a useful list of T-duality relations of twisted circle.

\newsec{Twisted Circles, ${\bf C}/{\bf Z}_n$ and Type 0 theory}

As we have explained in the introduction, in this paper we will study 
the evolution of stable D-branes in the decay of particular 
unstable closed string 
backgrounds. In this section we introduce these unstable backgrounds; 
Twisted Circles, ${\bf C}/{\bf Z}_n$ and type 0 theory. We will 
emphasize that two of these backgrounds, namely 
${\bf C}/{\bf Z}_n$ and type 0 may be realized as specific 
limits of the Twisted Circle theory.

\subsec{Twisted Circles}

Consider type II theory in flat ten-dimensional space
$(\vec x\in {\bf R}^{6+1},y\in {\bf R},z\in {\bf C})$, 
quotiented by a circle identification
accompanied by a rotation in an auxiliary plane
\eqn\twisid{
(\vec x,y,z) \sim (\vec x, y+2\pi R,e^{2\pi i\zeta}z).
}
We will refer to such a space (or the non-trivial three-dimensional part of
it) as a ``twisted circle'' with radius $R$. 
Note that the particular case $\zeta=0$ corresponds to the usual  
supersymmetric circle compactification, while $\zeta=1$ is a circle 
compactification accompanied by a $2 \pi$ rotation in a transverse plane,
i.e. a Scherk-Schwarz compactification.

Twisted sectors of \twisid\ are strings that wind around the $y$ direction
and are simultaneously constrained to rotate in an arc on the $z$ plane. 
As the squared length of such strings grows like $r^2$, generic
twisted sector states are localized to the origin of the $z$ plane. The
exact spectrum of type II theory on twisted circles is not difficult to 
derive (\refs{\tse,\TseytlinEI,\TseytlinZV}; see \refs{\TaUeo, \DaGuHeMi} 
for a brief review.) and confirms these qualitative expectations. 

For technical reasons we will restrict ourselves in this paper to the 
study of \twisid\ with a discrete set of twists; $\zeta={(n+1)/n}$ with 
$n$ odd i.e.
\eqn\twisidn{
(\vec x,y,z) \sim (\vec x, y+2\pi R,e^{2\pi i{n+1 \over n}}z).
}
Note that $2 \pi n \zeta=2 \pi(n+1) \approx 0$. Consequently 
it is best to think of \twisidn\ as a
${\bf Z}_n$ orbifold $({\bf C}\times {\bf S}^1)/{\bf Z}_n$
of a supersymmetric circle compactification 
\eqn\susycirc{
(\vec x,y,z) \sim (\vec x, y+2\pi n R, z),
}
of type II theory. States in the twisted sector of the ${\bf Z}_n$ 
orbifold are 
all localized on ${\bf C}$ while states in the untwisted sector 
are the usual
winding and Kaluza Klein modes on \susycirc\ and are delocalized. 

At large $R$ all twisted sector states are very massive and 
so  \twisidn\ is classically stable. On the other hand, 
it turns out that \twisidn\ is unstable at small $R$; more precisely, the 
spectrum includes at least one tachyon if and only if 
$2\apm (\f{n-1}{n})>R^2$; the 
mass of this state is $M^2=(R/\alpha')^2-2\alpha'(\f{n-1}{n})$. 
In this paper, 
we will be interested in \twisidn\ in its unstable regime. 

\subsec{${\bf C}/{\bf Z}_n$ from $R \to 0$}

In the limit 
$R \to 0$ \twisidn\ reduces to a simpler ${\bf Z}_n$ orbifold 
of \susycirc\ \TaUeo.
\eqn\twisidno{
(\vec x,y,z) \sim (\vec x, y, e^{2\pi i{n+1 \over n}}z);
}
i.e. the usual ${\bf C}/{\bf Z}_n$ orbifold in a plane transverse 
to the circle compactification. Note that the winding number of strings 
around the twisted circle in \twisidn\ is conserved; this conservation 
law modulo $n$ reduces to the `quantum symmetry' of 
the ${\bf Z}_n$ orbifold 
\twisidno. (Recall that the ${\bf Z}_n$ quantum symmetry of a ${\bf Z}_n$ 
orbifold \Vaq\ is generated by an element $h$ defined by the action 
\eqn\qs{h:|k\lb\to e^{\f{2\pi ik}{n}}|k\lb,} where 
$|k\lb$ is any state in the $k^{th}$ twisted sector).

In summary, the $R \to 0$, the IIA/B twisted circle model \twisidn\ 
reduces to IIA/B theory on ${\bf R}^7 \times {\bf C}/{\bf Z}_n$ times a
zero size circle, or (by T-duality) to IIB/A theory on ${\bf R}^8 \times 
{\bf C}/{\bf Z}_n$ \TaUeo.

\subsec{Type 0 from $n \to \infty$}

In the limit $n \to \infty$, IIA theory on \twisidn\ reduces to IIA theory
on a Scherk Schwarz circle, i.e. IIA theory modded out by 
$(-1)^{F_s} \times \sigma$ where $\sigma$ is the translation $y \to y+2 \pi
R$ and $F_s$ is the space time fermion number. This Scherk Schwarz
compactification of type II theory has been referred to as an 
`interpolating orbifold' \CoGu, as it interpolates between IIA theory on 
${\bf R}^{10}$ (as $R \to \infty$) and 0B theory on ${\bf R}^{10}$ 
(as $R \to 0$). (See Appendix A for more details). Consequently, one may
obtain type 0B/A theory from the twisted circle of IIA/B theory by first 
taking  $n \to \infty$ and then taking $R \to 0$ \DaGuHeMi .

\newsec{Closed String Tachyon Condensation in Twisted Circles}

The backgrounds described in section 2 are all unstable; their 
spectrum includes at least one closed string tachyon. In this section, first
we review the tachyon condensation\foot{Recall ${\bf C}/{\bf Z}_n$ 
and type 0 theories may be obtained as 
limits of twisted circle theories.} of twisted circle theories \DaGuHeMi . 
After that we derive and study the smooth 
`classical' Euclidean 4-geometry of the corresponding tachyon condensation 
process in detail. The intuition we gain from this study will
prove useful to us in section 5 below. 

\subsec{Decay of Twisted Circles in Gauged Linear Sigma Model Description}

In this subsection we will review a construction of a renormalization 
group flow that describes the tachyon condensation of twisted circles 
\DaGuHeMi. The construction of this renormalization group flow uses 
${\cal N}=(2,2)$
gauged linear sigma models \WittenYC \HoKa\
(see \DaGuHeMi\ and references therein for 
all details) and is similar to Vafa's construction \Va\ of a 
flow for ${\bf C/Z}_n$ (reviewed in detail in Appendix B).

Consider a gauged linear sigma model that contains two charged 
superfields $\Ph_1$ and $\Ph_{-n}$ that transform under $U(1)$ gauge 
transformations as 
\eqn\ft{
\Ph_1 \to e^{i\a} \Ph_1,\qquad
\Ph_{-n} \r e^{-in \a} \Ph_{-n},
}
together with an axionic field $P=P_1+iP_2$ 
whose gauge transformation (see \HoKa\ for general arguments) is
\eqn\ft{
P \to P + i \alpha.}
Note that 
$P_1$, the real part of $P$, effectively plays the role of a dynamical FI 
term. As in B.2, the low energy dynamics is described by a quantum corrected 
sigma model on the supersymmetric manifold of zero-energy configurations 
satisfying (D-term condition)
\eqn\zet{-\f{D}{e^2}=
|\ph_1|^2-n|\ph_{-n}|^2+k P_1=0,
}
modulo gauge equivalences, where $k$ is the coefficient of the kinetic 
term of the $P$ field (see \DaGuHeMi ) \foot{ More precisely, we study the 
gauged linear sigma model modded out by a discrete non anomalous
R symmetry (see B.3). This modding out imposes the type II GSO 
projection on the low energy sigma model.}.

At large values of $P_1$ (UV region), we use the
gauge freedom to set $\ph_{-n}$ to be real and positive, and then use
\zet\ to solve for $\ph_{-n}$ as $\ph_{-n} = \sqrt{(|\ph_1|^2 +k P_1)/n}$.
Plugging this into the classical action (see Appendix C for details)
we obtain the flat sigma model on the complex plane ${\bf C}$ 
(parameterized by $\ph_1$) times a
cylinder (parameterized by $P$) orbifolded by the unfixed ${\bf Z}_n$ 
gauge symmetry
\eqn\unfixexgt{
\ph_1 \r e^{{2 \pi i/ n}} \ph_1, ~~~P_2 \r P_2+{2 \pi i \over n}.
}
Consequently, at large $P_1$, dynamics reduces to that of 
a sigma model on the
twisted circle theory $({\bf C}\times {\bf S}^1)/{\bf Z}_n$
with a circle of radius 
$R_{UV}={\sqrt{\al k} \over n}$. Its metric is given by
\eqn\twim{ds^2= (d\rho)^2+\rho^2(d\theta)^2
+\f{k}{2}(dP_1)^2+\f{k}{2}(dP_2)^2,} where $\theta$ is the angle
of\foot{In Appendix C we denote this angle by $\theta_1$ just for notational
convenience. Both are the same.} 
$\ph_1$ and we defined $\rho=|\ph_1|$.

In \DaGuHeMi\ it is argued that $P_1$ may be regarded as a 
Liouville or scale 
direction, with $P_1 \r \infty$ corresponding to the UV.  
Consequently, the evolution of our target space in the $P_1$ 
mimics the flow of the energy scale from UV to IR as the change of $P_1$ 
from $P_1=\infty$ to $P_1=-\infty$. In this sense the coordinate 
$-P_1$ can be regarded as 
the real time $t$ after the Wick rotation. Consequently, the 
end point of tachyon condensation in the twisted circle theory is 
captured by the behavior of the sigma model as $ P_1 \r -\infty$. In this 
limit we obtain the cylindrical flat space ${\bf C}\times {\bf S^1}$ 
with radius $R_{IR}=\sqrt{\al k}$
\eqn\flatm{ ds^2= d\rho'^2+n^2\rho'^2 (d\theta)^2
+\f{k}{2}(dP_2-d\theta)^2,}
where we defined $\rho'=|\ph_{-n}|$.
Thus we conclude that the twisted circle
theory decays into a supersymmetric circle compactification of flat space.

Now consider the limit $k\to 0$; the twisted circle reduces to the orbifold 
${\bf C}/{\bf Z}_{n}\times {\bf S^1}$, where the radius $R'$ of this 
${\bf S}^1$
is $n$ times the radius $R$ of the twisted circle background \TaUeo. 
According to 
\flatm, the end point of this decay is ${\bf C}\times {\bf S^1}$, where 
the radius of ${\bf S}^1$ is given by $n R = R'$. Consequently, 
the decay of 
${\bf C}/{\bf Z}_n \times {\bf S}^1$ 
replaces the orbifold by flat space, and leaves 
the ${\bf S}^1$ 
untouched, consistent with the conclusions of \AdPoSi\ \Va 
(for completeness we review tachyon condensation
in ${\bf C/ Z}_n$ in detail in Appendix B).

In the limit $n \to \infty$ of the twisted circle reduces to a 
${\bf Z}_2$ orbifold of type 0 theory on a circle of radius ${\apm \over R}$
(see previous section). If the description of tachyon condensation 
commutes with $n \to \infty$ (and it is not clear that it does), 
it follows that the end point of tachyon condensation of the ${\bf Z}_2$ 
orbifold of type 0A/B theory on an ${\bf S}^1$ (reviewed in subsection 2.3) 
is simply type IIB/A theory on ${\bf R}^{10}$.

\subsec{A Construction of the Interpolating Liouville Flow}

As we have reviewed in the previous subsection, the entire process of
tachyon condensation of the twisted circle theory is described by a
conformal field theory. The CFT that captures the details of this tachyon 
condensation process may be constructed via a two step procedure. In the 
first step we plug the D-term equation \zet\ into the GLSM action, 
fix the $U(1)$ gauge and then integrate out the (massive) gauge field which 
appears quadratically in the Lagrangian. These manipulations are easy to 
perform and lead, in the sigma model limit 
$e \to \infty$ to a ${\cal N}=(2,2)$ sigma model
on a smooth four dimensional target space\foot{
Here we again set the angle of the complex scalar field 
$\ph_{-n}$ to zero by gauge fixing. We can also employ
a gauge invariant expression without gauge fixing, which 
looks more complicated. Since the result does not change,
 we will use the former expression. We thank K.Hori very much 
for comments on this point.} 
 (see Appendix C for details)
\eqn\metrico{ds^2=  d \rho^2+ d \rho'^2
+\f{2}{k}(n\rho' d \rho' -\rho d\rho)^2 + 
 {{k \over 2} \rho^{2}d\tilde{\theta}_1^2
 +{k \over 2}\rho'^2d\tilde{\theta}_2^2
 +\rho^2\rho'^2 
 (nd\tilde{\theta}_1+d \tilde{\theta}_2)^2
 \over 
\rho^2 + n^2 \rho^{'2}+{k \over 2}  }.}
The coordinates that appear in \metrico\ are related to those in \twim\ and 
\flatm\ by $\tilde{\theta}_1\equiv \theta-P_2$ and 
$\tilde{\theta}_2=nP_2$. \unfixexgt\ implies that $\tilde{\theta}_{1,2}$ 
are periodic with periodicity $2\pi$, and that the radial parameters $\rho$ 
and $\rho'$ run from $0$ to $\infty$. 

In the next (and unsatisfyingly implicit) step of this construction we 
flow to the IR of this sigma model to obtain the desired CFT. 
Unfortunately, we have not been able to find an exact description of the 
resulting ${\cal N}=(2,2)$ SCFT. However, in this paper we will be interested 
only in qualitative aspects of of \metrico.
We will assume that, as in \HoKa, (and several earlier analysis of Calabi 
Yau constructions, see for instance \WittenYC\ ), the asymptotic geometry 
and topology contained in the exact SCFT match those of \metrico. In the
rest of this section we will study the geometry of \metrico\ 
(we refer to \metrico\ as the `classical' geometry of tachyon condensation).

\subsec{The Classical Geometry of Tachyon Condensation}

\fig{Geometry change under the closed 
string tachyon condensation. The bubble of flat space appears at the origin 
when $P_1=0$. Later it will expand and finally it covers the whole space.}
{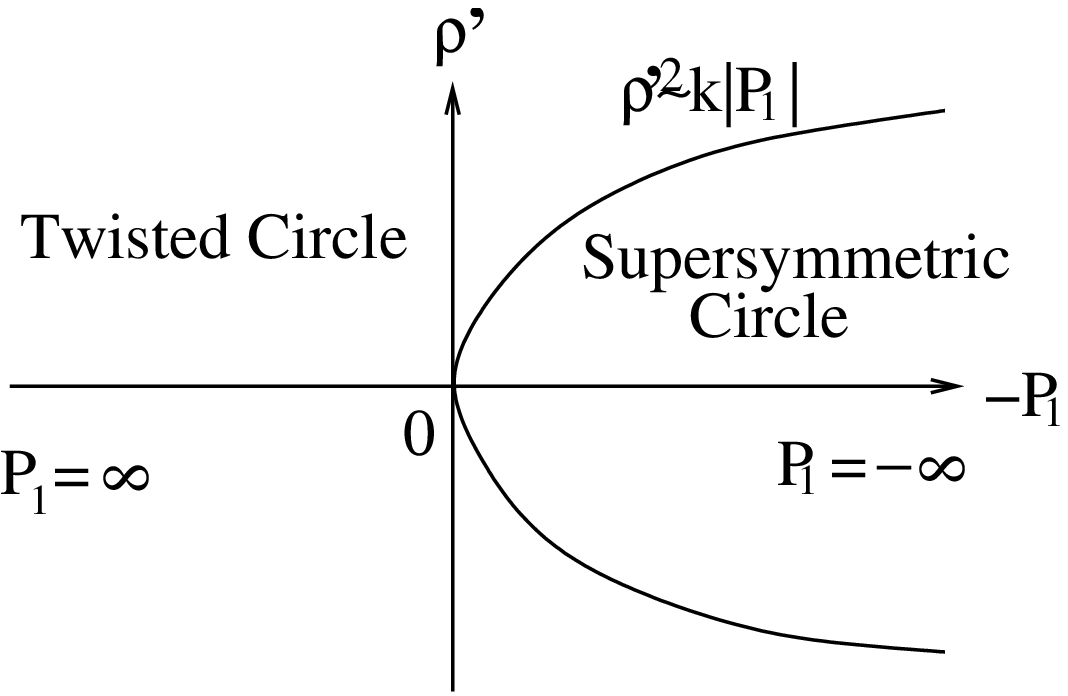}{3.0truein}

As we have argued above, \metrico\ describes a space that interpolates 
between the twisted circle geometry and flat space. It is easy to see 
this in detail. Rewriting the metric in terms of the coordinates 
$\rho$, $\theta$, $P_1$ and $P_2$; see around \ft - \flatm\ for notation)
we find that, in the UV limit ($P_1 \gg 0$) the interpolating space reduces 
to the
twisted circle \twim
\eqn\metricone{\eqalign{
ds^2&=(d\rho)^2+\f{k}{2}(dP_1)^2+\f{(\rho d\rho
+\f{k}{2}dP_1)^2}{n(\rho^2+kP_1)}\cr
&+\!\f{(n\rho^4+(nkP_1\!+\!\f{k}{2})\rho^2)(d\theta)^2
\!-\!k\rho^2(d\theta)(dP_2)\!+\!\f{k}{2}((n+1)\rho^2\!+\!nkP_1)(dP_2)^2}
{(n+1)\rho^2+nkP_1+\f{k}{2}}.\cr
& \approx (d\rho)^2+\rho^2(d\theta)^2
+\f{k}{2}(dP_1)^2+\f{k}{2}(dP_2)^2 ~~~~(P_1 \gg 0).\cr}}
On the other hand, rewriting 
the metric in terms of the coordinates $\rho'$, $\theta$, $P_1$ and 
$\tilde{\theta_1}$, we see that the interpolating space reduces to flat
space \flatm\ (i.e. supersymmetric circle) in the IR ($P_1 \ll 0$)
\eqn\metrictwo{\eqalign{
&ds^2=(d\rho')^2+\f{k}{2}(dP_1)^2+\f{(n\rho' d\rho'
+\f{k}{2}dP_1)^2}{n\rho'^2+kP_1}\cr
&+\!\f{\f{k}{2}(n\rho'^2\!+\!kP_1)(d\tilde{\theta}_1)^2
+n^2\rho'^2(n\rho'^2\!+\!kP_1)d\theta^2+\f{kn^2}{2}\rho'^2
(d\tilde{\theta}_1-d\theta)^2}
{n(n+1)\rho'^2+kP_1+\f{k}{2}}\cr
&\approx d\rho'^2+n^2\rho'^2 (d\theta)^2
+\f{k}{2}(d\tilde{\theta}_1)^2
.\cr}}

In greater detail, \metrico\ represents the decay of the twisted circle 
background via the  nucleation and growth of a bubble of supersymmetric 
$S^1$ compactification. This bubble is nucleated at 
`time' $P_1= 0$ (and at $\rho=\rho'=0$).  The radius $R_b$ of this 
nucleated bubble (see Fig.2) grows in time 
like $R_b\sim \sqrt{-kP_1}$ (this 
diffusive growth\foot{
In order to see this note that, in the far region $\rho'^2>> k|P_1|$ 
\metrico\ reduces to the twisted circle, while in the opposite limit 
$\rho'^2<< k|P_1|$ we get the flat space \flatm. $\rho'^2 \approx k|P_1|$
(see Fig.2) 
represents a transition between these regions. } 
matches the behavior observed in \AdPoSi, \GuHeMiSc\ ). These results
 precisely 
show the expected behavior of the tachyon condensation 
process (see Appendix
B.1 for a general discussion).

In order to understand the transition between the twisted circle and
supersymmetric geometry, it is useful to focus on the point at which 
the bubble of supersymmetric circle is first nucleated, i.e. the 
neighborhood of $P_1=\rho=\rho'=0$. \metrico\
reduces to
\eqn\metricflat{ds^2=d\rho^2+\rho^2 d\tilde{\theta}_1^2
+d\rho'^2+\rho'^2 d\tilde{\theta}_2^2,}
i.e. (smooth) polar coordinate of ${\bf R}^4$. 

It is interesting to trace the behavior of various cycles in the 
twisted circle, as
the space transforms itself from the twisted circle to flat space. The 
twisted circle ${\bf S}^1_{A}$ defined by (see Fig.3)
\eqn\defcc{{\bf S}^1_{A}:(\rho,\theta,P_2)=(0,s_1,s_1),\ \ 0\leq 
s_1 \leq 2\pi/n,}
which is  non contractible for $P_1 \gg 0$ (UV region), shrinks 
and disappears at $P_1=0$. This is
simply the angle circle $\tilde{\theta_2}$ of one of the two planes in 
\metricflat. This cycle becomes trivial for $P_1<0$.  
\fig{One cycles in twisted circle.}{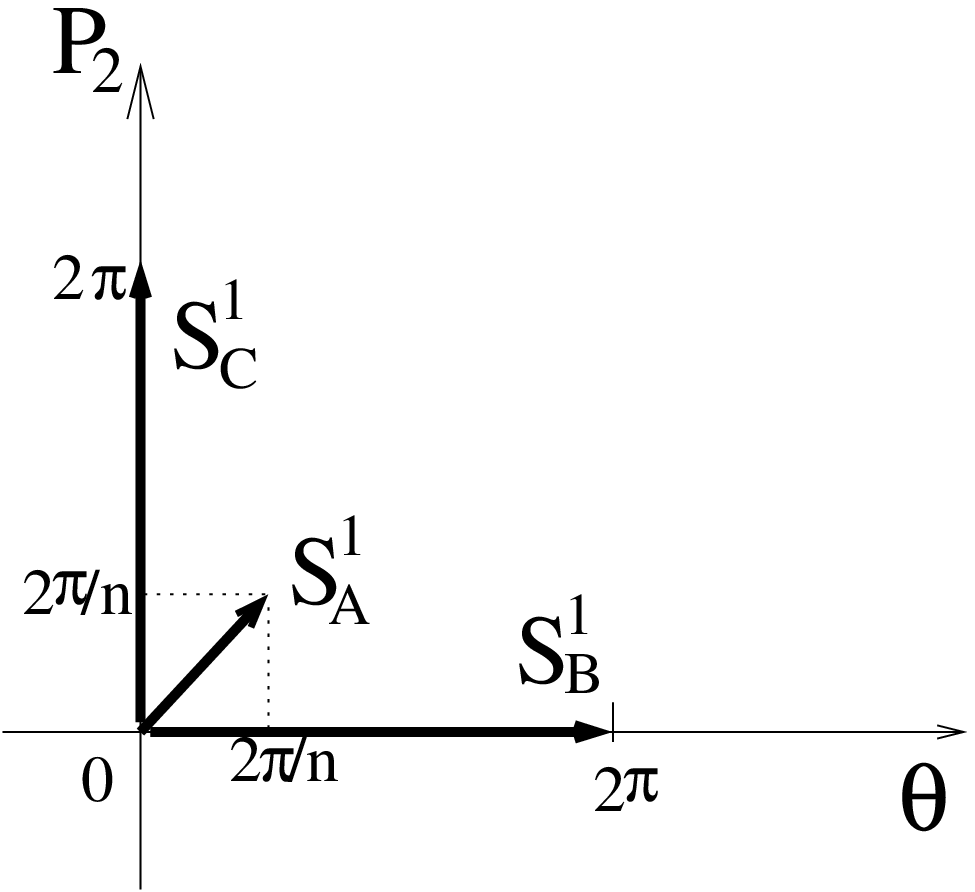}{2.5truein}

It is also interesting to track the evolution of the supersymmetric circle
${\bf S}^1_{B}$ defined by
\eqn\defthc{{\bf S}^1_{B}:(\rho,\theta,P_2)=(0,s_2,0),\ \ 
0\leq s_2 \leq 2\pi,}
back in `time', from the IR to the UV. This cycle (see Fig.3)
also vanishes at $P_1=0$; in fact it is simply the other angle 
${\tilde \theta}_1$ in \metricflat. 

In section 5 below we will be interested in the third cycle (see Fig.3)
\eqn\deftc{{\bf S}^1_{C}:(\rho,\theta,P_2)=(a,b,u),\ \
 0\leq u \leq 2\pi,\ a,b=constant}
which winds $n$ times around the twisted circle in the UV $P_1 \gg 0$ and 
winds once around the supersymmetric circle in the IR $P_1 \ll 0$. Note 
that this cycle vanishes at the point $\rho=P_1=0$. 

\newsec{D Branes on Twisted Circles, ${\bf C}/{\bf Z}_n$ and Type 0}

The spectrum of D-branes on Twisted circles has been previously derived
using a boundary state analysis \TaUet\ \DuMo. In this section
 we will review the results, and provide a spacetime 
 interpretation of these branes. 
We will also describe how these branes reduce, in the appropriate limits, 
to the branes of ${\bf C/Z}_n$ and type 0 theory.

\subsec{D Branes on Twisted Circles}

D branes on twisted circle backgrounds are of two sorts; those that live at
a point on the twisted circle, and those that wrap around the twisted
circle. 

A branes at a point on the twisted circle is best thought of
as an array of branes on the covering space (see, for instance \piljin). 
At low energies such branes behave much like their counterparts in type 
II theory on flat space.

D branes that wrap the twisted circle are more interesting.
Consider a 1-brane that wraps the twisted circle once.
As in the case of twisted sector strings, such a brane is confined to the 
origin of the $z$ plane. Classically, its mass is given simply by the usual 
D1-brane tension times the minimum length of the twisted circle 
$M={R \over g \apm}$. Of course such 1-branes actually appear in 
degenerate one parameter family of solutions labeled by the Wilson line 
$\theta=\int A \cdot dl \in [0, 2 \pi)$ along the brane.

As we have discussed above, the twisted circle may be regarded
as a ${\bf Z}_n$ orbifold of \susycirc, a circle of radius $nR$. 
The 1-brane of
the previous paragraph is `S-dual' to a twisted sector string
state of this orbifold; it is called a fractional brane. On the other hand
a 1-brane wrapping the twisted circle $n$ times is the analogue of an
untwisted sector string state, and is referred to as a bulk brane. 
We will now argue that the bulk 1-brane may be thought of as a 
collection of $n$ fractional 1-branes, with a distribution of Wilson
lines. Refer to \DoMo \DiDoGo\ for original 
literature on (fractional) D-branes 
in supersymmetric orbifolds (e.g. ${\bf C}^2/{\bf Z}_n$).  

A collection of $n$ fractional 1-branes is described, at low energies, 
by a $1+1$
dimensional $U(N)$ gauge theory, with adjoint matter fields $Z$, $\psi$
and $\phi$, where $Z$ represents the position of the brane in the $z$ plane, 
$\psi$ is the corresponding fermion, and $\phi$ represents all other 
massless fields 
(the gauge field, fluctuations in $R^7$ plus corresponding fermions). In
the vacuum describing $n$ separate 1-branes, these adjoint fields obey the 
boundary conditions\foot{
In the full string theory
\eqn\bcf{ \chi(2 \pi R)= e^{{2 \pi i k(n+1) \over n}}C \chi(0) C^{-1}}
where $k=n_z-n_{\bar z}+\half(n_{\psi}-n_{\bar \psi})$, where
$n_z, n_{\bar z}, n_{\psi}, n_{\bar \psi}$ respectively represent the
number of $Z$, $\bar Z$, $\psi$ and $\bar \psi$ open string oscillators
excited in producing the open string field $\chi$. }
\eqn\bc{\eqalign{ Z(2 \pi R)=& e^{{2 \pi i (n+1) \over n}}S Z(0) S^{-1}, \cr
             \psi(2 \pi R)=& e^{{2 \pi i (n+1) \over 2n }}S 
             \psi(0) S^{-1},\cr
              \phi(2 \pi R)=& S \phi(0) S^{-1}. \cr }}
where $S$ is a gluing $U(n)$ matrix representing the Wilson line on the 
brane. Different values of $S$ represent different classical vacua of
this system. 
In particular $S=I$ represents a configuration of $n$  
strings, 
independently winding the twisted circle. On the other hand, when 
$S$ is chosen as  the shift matrix, the $n$ 1-branes are effectively 
strung together into a bulk brane; a single brane winding the circle $n$ 
times. As the shift matrix has eigenvalues
\eqn\eigenvalues{e^{2 \pi i k \over n},~~~ k= 1 \ldots n.}
we conclude that bulk 1-brane is simply a collection of 
$n$ fractional 1-branes with a relative collection of Wilson lines given by 
\eigenvalues. 

Wrapped 3-branes on the twisted circle are rather similar to
1-branes. Consider a single space-filling brane in the twisted circle 
background. Gauge fields on a collection of 
$n$ such D3-brane, obey the boundary conditions
\eqn\bco{\eqalign{ Z(y+2 \pi R, e^{{2 \pi i \over n}} z)=
& e^{{2 \pi i (n+1) \over n}}C Z(y, z) C^{-1} \cr
             \psi(y+2 \pi R, e^{{2 \pi i \over n}} z)=&
e^{{2 \pi i (n+1) \over 2n }} C \psi(y,z) C^{-1} \cr
              \phi(y+2 \pi R, e^{{2 \pi i \over n}} z)=&
C \phi(y,z) C^{-1}\cr }}
where the fields have the same meaning as in \bc\ and $C$ is the Wilson
Line\foot{The pure gauge configuration corresponding to (for instance) 
a U(1) Wilson lines is $(A_y, A_z, A_{\bar z})=({\theta \over 2 \pi R}, 0,
0)$.}.
When $C$ is the shift matrix, the $n$ `fractional' 3-branes are strung 
together into a single 3-brane winding the circle $n$ times, i.e. a bulk 
3-brane.  

Finally, it is easy to verify the absence of 2-branes that wrap the 
twisted circle $m<n$ times, as no such brane is flat in the 
covering space. It is, of course, not difficult to construct bulk
2-branes which wrap the twisted circle a multiple of $n$ times.

\subsec{D-Branes on ${\bf C}/{\bf Z}_n$}

We now study the D-branes described in the previous section, in
the limit $R \to 0$. According to subsection 2.2, these branes must reduce 
to D-branes of ${\bf C/Z}_n \times {\bf S}^1$. In this subsection we will
explain how this works in detail. 

Consider a unit fractional 1-brane of the twisted circle theory, in the
limit $R \to 0$. Such a brane wraps once around the twisted circle (of
radius $R$), and so in some sense, wraps ${1 \over n}$ of the larger
${\bf S}^1$ factor in ${\bf C/Z}_n \times {\bf S}^1$, see subsection 2.2. 
We will now argue
that there are $n$ different flavors of this 1-brane. 

Consider two unit fractional 1-branes of the twisted circle theory, 
with relative Wilson line $\theta$. Open string modes stretching 
between these two branes are charged under the relative gauge field, and 
so obey boundary conditions \bcf\ with $C \chi(0) C^{-1} = e^{i \theta} 
\chi(0)$. When $\theta =0$, only open string modes with $k=0$ mod $n$ 
($k$ is defined in \bcf) survive the $R \to 0$ limit.  At generic values of 
$\theta$ all modes are infinitely massive. However, in the neighborhood of 
$2 \pi \theta={m \over n}$, a new set of modes become 
massless, namely those 
with $k=m$ mod $n$. Consequently, there are $n$ species of 1-branes that 
can communicate with the $\theta=0$ brane; branes with 
$\theta ={2 m \pi \over n}+ \CO(R)$ where $m$ runs between $0$ and $n-1$. 

These results are easier to visualize upon T-dualizing along 
the ${\bf S}^1$ 
factor of ${\bf C/Z}_n \times {\bf S}^1$. The stretched one branes of
the previous paragraphs turn approximately into usual 0-branes on a circle 
of size $R'={ \apm \over n R}$. Turning on a relative Wilson line $\theta$
corresponds to separating the branes by $\Delta y= \theta \apm / 
R={\theta n R'}$. In taking $\theta$ from 0 to $2 \pi/n$ we take 
one of the 0-branes once around the circle; that process converts it into 
a zero brane of another variety! Repeating this operation produces yet
another zero brane, but, upon repeating the process $n$ times we revert to
the initial state. 

Below we will find use for an operator $Q$ 
whose action 
is to transport a zero brane once around the circle (in T-dual language 
$Q$ shifts the Wilson line by $e^{{i 2 \pi \over n}}$). 
If the $n$ distinct
0-branes are denoted by $|i\rangle$, then clearly $Q |i 
\rangle=|i+1 \rangle$.

In ${\bf C/Z}_n \times {\bf R}^8$ 
(the formal result of taking $R$ to zero on the 
twisted circle theory) the $n$ distinct species of 0-branes are completely 
distinct; they cannot be continuously deformed into each 
other; restated ${\bf C}/{\bf Z}_n$ has $n$ distinct fractional 
0-branes. However the operator $Q$ of the previous paragraph is well
defined as $R \to 0$ and reduces to the generator $h$ 
of the quantum symmetry \qs\
of ${\bf C}/{\bf Z}_n$ in the limit\foot{This is easily seen; $Q$ generates 
discrete translations, and so counts discrete moment or discrete windings
in the T-dual picture.} and so we conclude that the fractional 0-branes are
related to each other by the action of the quantum symmetry. It is well 
known that ${\bf C}/{\bf Z}_n$ does indeed have $n$ fractional 0-branes with
these properties. Of course the discussion of
this section may be mimicked to derive similar results for 
2-branes of ${\bf C}/{\bf Z}_n$.

To summarize, we have found that the familiar fractional 0-branes (and
2-branes\foot{Note that a single D2-brane in ${\bf C}/{\bf Z}_n$ should be
treated as a fractional D2-brane from the group theoretical viewpoint 
of \DoMo \DiDoGo (belonging to one of $n$ irreducible representations of 
${\bf Z}_n$). However, its tension
is not fractional but usual value. Thus
we will call it just a D2-brane in this paper.}) 
of ${\bf C}/{\bf Z}_n$ have a simple geometric
interpretation upon `blowing up' the ${\bf C}/{\bf Z}_n$ singularity 
into a twisted 
circle; they correspond to 1-branes (resp 3-branes) wrapping the twisted 
circle. This geometrical intuition will help us understand the fate of 
fractional 0-branes during tachyon condensation, later in this paper.

\subsec{D-Branes in Type 0 Theory}

In the limit $n \to \infty$, the IIA/B twisted circle turns into a 
Scherk-Schwarz 
circle of radius $R$. As we have reviewed in Appendix A, the T-dual 
of the Scherk-Schwarz compactification of type IIA/B theory is 
the ${\bf Z}_2$
orbifold of 0B/A on $S^1$ of radius ${\apm \over R}$, where the 
${\bf Z}_2$ is generated by $(-1)^{F_L} \times \sigma'$ and  $\sigma'$ is 
a shift across half the circumference of the circle. Of course, as $R \to
0$ this theory reduces locally to 0B/A in flat space. 

In the limit $n \to \infty$ the fractional 1-brane of the twisted circle is
simply a D1-brane wrapping the Scherk-Schwarz circle. We will now
investigate what happens to this D1-brane upon T-dualizing along this
circle. 

Recall that flat space string theories with $N=(1,1)$ world sheet
supersymmetry have two 0-brane boundary states with 
different spin structures
denoted by $|+\rangle $ and $|-\rangle$ (e.g. see \bega ). 
These boundary states obey 
\eqn\pmsta{\eqalign{&(-1)^{F_L} |+\rangle =-|-\rangle, \cr 
                    &(-1)^{F_R} |+\rangle =-|-\rangle, \cr
		    &(-1)^{F_L} |-\rangle =-|+\rangle, \cr 
                    &(-1)^{F_R} |-\rangle =-|+\rangle}}
{}From \pmsta\ we conclude that 
\item{1.} The only state invariant under $(-1)^{F_L}$ and $(-1)^{F_R}$ is
$$|+\rangle - |-\rangle ;$$ 
this is the 0-brane of type IIA theory.
\item{2.} Any linear combination of 
$$|+\rangle \   {\rm and} \ |-\rangle ,$$
is invariant under 
$(-1)^{F_L+F_R}$; these are the so called electric and  magnetic 0-branes
of type 0A theory.  
\item{3.} 
$$|+ (x)\rangle ~~- ~~|-(x+\pi R) \rangle ,$$ (where the $(x)$ in 
brackets refers to the position of the brane on the circle) is the only 
combination of 0A zero branes invariant under the ${\bf Z}_2$ orbifold of 
0A on a circle of size $ 2 \pi R$, described earlier in this section.

With these facts in mind, it is natural to guess, and easy to verify, that 
the T-dual of a 1-brane wrapping the thermal IIB circle is the 
a configuration of electric and magnetic 0-branes described in item 3 
above. 

\subsec{Dirac Quantization}

In this subsection we will study the charges of the fractional branes
described in the previous subsection, and explain how they are consistent 
with Dirac quantization. This subsection is a side branch away from the 
main line of development of this paper. The reader eager to turn to the
fate of D-branes under closed string tachyon condensation can skip to the
next section.

Consider an action that describes the interaction of 
3-form and 7-form field strengths with 1-branes and 5-branes
\eqn\acoftf{\eqalign{S&=C\int d^{10}x F_3^2+ e\int A_2 + g\int A_6 \cr
&=C\int d^{10}x F_7^2+ e\int A_2 + g\int A_6 ,}}
where $F_3=dA_2$, $F_7=d A_6$ and $F_7=*F_3$ and $A_2$ (resp $A_6$) is 
integrated over the world volumes of the 1-branes (resp 5-branes).
Dirac quantization imposes the constraint 
\eqn\dccond{{eg \over C} = {2 n \pi},}
on the charges in \acoftf. In particular, the minimally charged 1-brane
(charge $e_0$) and 5-brane (charge $g_0$) obey 
\eqn\mincp{{e_0 g_0 \over C}=2 \pi.}

As an exercise that will be useful later in this subsection, consider 
the dimensional reduction of \acoftf\ (with minimally charged 1-branes
and 5-branes) on ${\bf T}^2 \times{\bf  S}^1$ 
where the torus has volume $V_2$ and the 
circle has radius $R$. Let $A_1$ denote the reduction of $A_2$ along $S^1$, 
let $A_4$ denote the reduction of $A_6$ along $T^2$. 
The dimensionally reduced
action that governing the interaction of particles (10 dimensional strings 
wrapping $S^1$) and 3-branes (10 dimensional 5-branes wrapping the $T^2$) 
with $A_1$ and $A_4$ is
\eqn\dimredact{S=C (2 \pi R) V_2\int d^{7}x F_2^2+ e_0(2 \pi R)
\int A_1 + g_0V_2\int A_4.}
Note, of course, that, as a consequence of \mincp,  the charges in 
\dimredact\ automatically saturate the 7-dimensional Dirac Quantization 
constraint 
${(e_0 2 \pi R) (g_0V_2) \over C (2 \pi R) V_2   }=2 \pi$.

We now adapt the exercise of the previous paragraph to a compactification
$({\bf T}^2  \times {\bf S}^1)/ {\bf Z}_n$ 
of the twisted circle background 
$({\bf C}  \times {\bf S}^1)/ {\bf Z}_n$. 
For clarity consider a specific 
case; let $n=3$ and let the $z$ plane in \twisidn\ be an `equilateral 
2-torus' (the complex plane modded out by translation vectors at an 
angle ${2 \pi \over 3}$ to each other) of volume $V_2$. The 3-volume of
this twisted circle background is $2 \pi R V_2$. The length of the smallest
1-cycle along the twisted circle is $2 \pi R$. The volume of the minimal 
dual 2-cycle is $V_2$ for any nonzero $R$ and, as in the previous
paragraph, the electric and magnetic charges of the dimensionally 
reduced 7 dimensional theory automatically obey the Dirac quantization 
condition\foot{These arguments explain the phenomenological 
observation noted in \TaUet\ that 
we cannot find any D-brane in the Twisted Circle which is 
localized in the circle direction and 
which
becomes a fractional D-brane in the $R\to 0$ limit.
Such a brane, if exists, will break the Dirac quantization rule.}
. 

We now turn to the $R \to 0$ limit of this situation. As we have explained
in subsection 2.2, the twisted circle reduces to 
${\bf T}^2 / {\bf Z}_3 \times {\bf S}^1$ where the radius of the 
${\bf S}^1$ is 
$3 R$. Note that the minimal volume 2-cycle along the $T^2$ in this
orbifold has volume $V_3/3$. Further, although the orbifold 
circle has length 
$3R$, the effective charge of the wrapped 1-brane (which reduces to the
fractional 0-brane as $R \to 0$), continues to be $e_0 2 \pi R$ (that is
why this object is a fractional rather than bulk brane) and the 
effective 7-dimensional action governing the electromagnetic interactions
between the 2-brane (i.e. `fractional' 
2-brane wrapping a cycle of volume $V_2/3$)
and the fractional 0-brane is 
\eqn\dimredact{S=C (2 \pi R) {V_2}\int d^{7}x F_2^2+ 
{e_0 {2 \pi R }}
\int A_1 + g_0{V_2 \over 3}\int A_4.}
At first sight \dimredact\ appears to violate 
Dirac Quantization (as ${(e_0 {2 \pi R }) (g_0{V_2 \over 3}) 
\over {C (2 \pi R) V_2 }   }={2 \pi \over 3} \notin 2 \pi {\bf Z}$).  
The resolution to this puzzle is rather instructive. The 2-branes 
and fractional 1-branes of ${\bf T}^2 / {\bf Z}_3 \times {\bf S}^1$ 
are certainly
charged under untwisted RR sector potentials. However they are 
equally charged under all twisted sector RR 2 form potentials 
(which are massless for ${\bf T}^2 / {\bf Z}_3 \times {\bf S}^1$
 but not for the 
twisted circle theory at any nonzero $R$). The true phase in taking a 
fractional 0-brane around the Dirac String attached to the
2-brane is the three times the naive result from \dimredact, restoring
consistency.

\newsec{D-branes Under Tachyon Condensation}

In this section we turn to the main topic of this paper; an 
investigation of the fate of D-branes under tachyon condensation. Here 
we consider the evolution of D-branes only under bulk flows that 
lead to a supersymmetric end point; we postpone the discussion 
of fine tuned flows to the next section. We will 
first state our conjectures, and then proceed, in the rest of this section,
to provide evidences for them from two different viewpoints. 

{\bf Twisted circles} :  Any brane located at a 
point on the twisted circle evolves into the corresponding brane at a point 
on the supersymmetric circle. 3-branes that wrap the twisted circle once 
unambiguously evolve into usual 3-branes that wrap the supersymmetric circle 
once. All fractional 1-branes disappear in the process of tachyon 
condensation. A bulk 1-brane (which wraps the twisted circle $n$ times) 
either disappears (this is possible only near the origin) 
or evolves into a bulk 1-brane that wraps the 
supersymmetric circle once.

{\bf Orbifold ${\bf C}/{\bf Z}_n$}: All $n$ varieties of space filling 
2-branes evolve into the usual space filling 2-branes in flat space. 
Every fractional 0-brane disappears in the process of tachyon condensation.
A bulk 0-brane either disappears 
(again this is possible only near the fixed point) or 
evolves into a usual 0-brane in flat space.

{\bf ${\bf Z}_2$ Orbifold of type 0A on ${\bf S}^1$}: 
It should be emphasized that all results about type 0 theory are obtained 
upon taking a potentially dangerous $n \to \infty$ limit. 
With this caveat in mind, we find the following results. 
A pair of electric and magnetic D(p+1)-branes wrapping 
the circle evolve into a single Dp-brane in IIB theory. 
A pair of diametrically separated electric 
and magnetic 0-branes disappear in the process of tachyon condensation. 
On the other hand, diametrically opposed 0A 2-branes 4-branes and 6-branes
evolve respectively into single IIB 3-branes, 5-branes and 7-branes.

We will now proceed to provide evidence for these conjectures. 
The rest of this section is organized as follows. In subsection 5.1 we 
explain how the results for twisted circle theories follow from topological 
properties of the interpolating geometry presented in the previous
section. Results for ${\bf C/Z}_n$ and type 0 theory follow 
from the appropriate limits (see section 2 and 3) of the twisted circle
theory. In subsection 5.2 we comment on how our results are consistent with 
charge conservation. In subsection 5.3 and 5.4, respectively, we provide
independent evidence for our conjectures by studying the boundary condition 
of world-sheet under the RG-flow in GLSMs. In appendix D we find further 
confirmation of our results by an analysis of D-branes in the dual LG models.

\subsec{Geometrical Analysis}

Let us first consider the behavior of a fractional D1-brane
under closed string tachyon condensation in the twisted circle theory
from geometrical viewpoint. As we have explained in the previous
section, a D1-brane that wraps the twisted circle once is called a 
fractional D1-brane. In order to understand the fate of this 
fractional brane under tachyon condensation, we can follow the
evolution of this cycle as tachyon condensation proceeds.
The geometry of the GLSM model was discussed in detail in section 3; 
as we argued there, the twisted circle ${\bf S}^1_A$ vanishes at 
$P_1=\rho=0$
and cannot be extended beyond this point. Indeed, near that special
point this cycle is simply an angle in a 2-plane in ${\bf R}^4$.
Consequently a D1-brane wrapping the twisted cycle
circle simply self annihilates at $P_1=\rho=0$. The induced metric 
on the world volume of such an annihilating brane (along one possible 
trajectory) is given by the everywhere smooth cigar (see Fig.4)
\eqn\tcam{ds^2=\f{k}{4nP_1}(dP_1)^2+knP_1(ds_1)^2
=(dr)^2+n^2r^2(ds_1)^2\ \ (r\equiv \sqrt{P_1 k/n}),} 
\fig{The smooth cigar and wrapped D1-brane.}{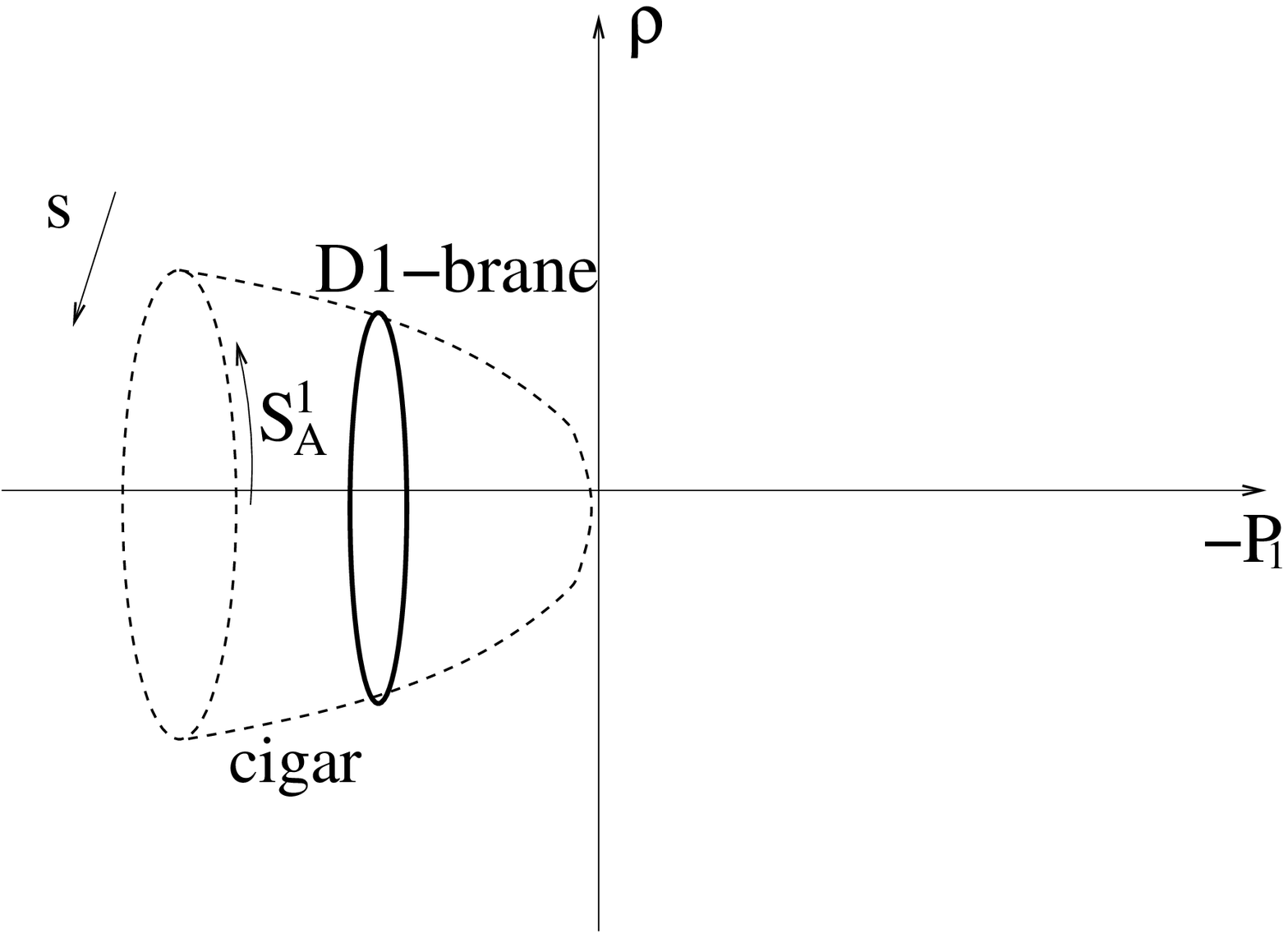}{3truein}

In contrast consider a bulk D1-brane. Such a brane wraps the twisted
circle $n$ times at early `times', and so may be thought of as a
brane wrapping the cycle ${\bf S}^1_{C}$ described in section 3
above. Notice that  ${\bf S}^1_{C}$ is nontrivial in the UV ($P_1 \gg
0$) as well as in the IR. However this cycle degenerates at a point 
along this RG flow, and so fall in the class described in item 3. in the 
introduction. Correspondingly, we conjecture that a bulk 0-brane can 
either disappear or evolve into an ordinary brane wrapping the 
supersymmetric circle.\foot{
Note also that ${\bf S}^1_{C}$ always has a non-zero radius except
$P_1=0$. Moreover, we can slightly shift the position of ${\bf S}^1_{C}$
from the origin and obtain a non-contractible one cycle. In such a situation
we would expect the bulk brane to survive the process of 
tachyon condensation.}

On a slightly different note, observe that the supersymmetric cycle 
${\bf S}^1_{B}$ describes the creation of a bulk zero brane 
(of the supersymmetric background), correspondingly we expect to find 
an RG flow that describes bulk 0-branes popping out of nothing. 

It is also possible to 
extend our considerations to D3-branes. Since the
world-volume for such a  brane fills all of space in the UV, it will  
continue to do so in the IR; consequently the 3-brane of 
the twisted circle evolves into a usual D3-brane in flat space after
the closed string tachyon condensation. As we have explained in section
4.1, twisted circle D3-branes appear in a one parameter family of solutions
labeled by the Wilson line around the twisted circle. It is not difficult 
to trace the evolution of this Wilson line into the IR; see 
subsection 5.3 below.

Finally we consider the behavior of D0-branes and D2-branes in the orbifold 
${\bf C}/{\bf Z}_n$. This is obtained by the small radius limit
$R\to 0$ as we have explained\foot{ We believe this limit 
is smooth as evidenced from e.g. the structure of GLSM \DaGuHeMi\ and the
calculation of partition function \TaUeo.}
 in section 3. 
Thus we conclude that a fractional D0-brane will
disappear after the tachyon condensation; a bulk D0-brane
either disappears or evolves into a bulk D0-brane in flat space, while 
a bulk D2-brane evolves into its supersymmetric counterpart (we will 
explain in subsection 5.3, all $n$ ${\bf C/Z}_n$ D2-branes have the same 
end point under
tachyon condensation). The latter fact for a bulk D0-brane 
is consistent with the results 
in \AdPoSi. These conclusions will be confirmed by investigating 
of the boundary interactions in the later subsections.

\subsec{Charge Conservation}

The `disappearance' of a D-brane under tachyon condensation naively
appears to be in conflict with charge conservation. Our geometrical
description of this process makes it clear that, at least in the 
Euclidean version of this process, this `paradox' is spurious. 

As we have described above, the world volume of the `disappearing' 
fractional D1-brane is the everywhere smooth `cigar' \tcam. 
This Euclidean D1-brane couples to the RR-2 form field
$F_{\mu\nu}$ through the everywhere conserved current 
\eqn\cur{\sqrt{G}J^{\mu\nu}=\int d^2\sigma \f{\de\xi^\mu}{\de\sigma^{\ap}}
\f{\de\xi^\nu}{\de\sigma^{\beta}}\ep^{\ap\beta}\prod_{\lambda}
\delta(\xi^{\lambda}(\sigma)-x^{\lambda}); }
certainly charge conservation is nowhere violated. 

The disappearance of D-brane charge can be made to appear more puzzling 
upon dimensionally reducing along the twisted circle. The fractional 
D1-brane is charged under the gauge field obtained from the Kaluza-Klein 
reduction of $F_{\mu\nu}$; this charge disappears at `time' $P_1=0$. 
Of course `charge conservation is violated' only when the twisted circle 
shrinks to zero size; precisely at this point the dimensionally
reduced description degenerates (for instance the 9 dimensional coupling 
blows up). The correct description is 10 dimensional, and
manifestly charge conserving.

Of course all considerations that apply to D-strings wrapping the twisted 
circle also apply to ordinary strings wrapping the twisted circle. 
However, as we have explained behind, such strings belong to the
twisted sector. Thus we conclude that twisted sector charges disappear
in the process of tachyon condensation.\foot{In \MaMo\ another
mechanism of the disappearance of the twisted RR-field
was proposed in non-supersymmetric orbifold theories. It would be interesting
to study the relation between this and our geometrical result.}

Recall that the reduction of the twisted circle to the
orbifold is most conveniently carried out after a T-duality;
consequently, it is instructive to examine the issue of charge
conservation after T-dualizing. In \HoKa\ 
Hori and Kapustin have demonstrated that
the T-duals of gauge linear sigma models, including 
2d Black hole \wittenbh\ and our model defined by \ft,
lead to super potentials that break
translational invariance along the dual circle. Roughly speaking, the 
potential may be thought of as a condensate (or plasma) of gravitons
with positive and negative momentum, 
explaining the lack of conservation of winding or
twisted strings prior to T-dualizing. It seems possible that this T-dual 
geometry should also be thought of as possessing a condensate of 0-branes
and anti 0-branes, explaining the violation of 0-brane charge conservation.

\subsec{GLSM analysis of ${\bf C}/{\bf Z}_n$ Boundary Flows}

In the previous subsections we have analyzed the evolution of the D-brane 
spectrum in twisted circle theories (and hence ${\bf C/Z}_n$) utilizing 
approximate geometry for Euclidean twisted circle tachyon condensation. 
In this subsection we will check some of those results using more rigorous
methods. Recall that twisted circle and ${\bf C/Z}_n$ 
world sheet theories are 
described by the UV of a GLSM (see section 3). In this section we will
explicitly construct the GLSM boundary conditions that correspond 
to certain D-branes in these backgrounds. We then analyze the evolution of 
these boundary conditions under world sheet RG flow; the fate of the 
D-brane in question is determined by the IR limit of these 
boundary conditions.  

The analysis of this section makes crucial use of the ${\cal N}=2$ 
supersymmetry in the world-sheet GLSM theory. Several authors (see 
for instance \HoIqVa \Gob \Ho \HeMc \Goc \HeKaLaMc; see also \OoOzYi\ 
for original arguments on boundary states) 
have studied GLSMs on world sheets with boundaries, and determined
conditions to ensure that the boundary conditions preserve 
${\cal N}=2$ B-type 
supersymmetry, and we will utilize their results below.

{\bf D2-branes}

As we have reviewed above, ${\bf C/Z}_n$ possesses $n$ different
(fractional) 2-branes. We first consider the simplest of these 2-branes 
(denoted by $|0\rangle$ in section 4 above). In the GLSM,
this 2-brane corresponds to
the following boundary condition \Ho\ assuming $\theta=0$
 (see Appendix B for details of this GLSM)
\eqn\bcdtt{{\cal D}_+ \Phi={\cal D}_- \Phi,\ \  \Sigma=\bar{\Sigma}, ~~~
v_{01}=0,}
where $\Phi$ denotes the chiral superfields whose lowest components 
are $\phi_1$ and $\phi_n$, $\Sigma$ is the twisted chiral (vector)
multiplet, and $v_{01}$ is the field strength. These boundary conditions 
are not modified by RG flow\foot{From the sigma point of view 
this follows as
the beta function for the spacetime gauge field vanishes when evaluated on 
the configuration $A_{\mu}=0$. This result also follows from the 
Landau Ginzburg analysis, see Appendix D.}.

We now turn to $|m\rangle$, the remaining $n-1$ types of 
D2-branes $(m\neq 0)$ in 
${\bf C/Z}_n$. Recall that $|m\rangle = h^m |0\rangle$ where $h$ generates
the ${\bf Z}_n$ quantum symmetry of ${\bf C/Z}_n$. 
In fact, the quantum symmetry acts
very simply on the (UV of the) GLSM; it effectively shifts\foot{
This fact can be easily understood in the Landau Ginzburg dual
picture (see Appendix D). The shift $h:U\to U+\f{2\pi i}{n}$ of 
quantum symmetry is equivalent to the shift of imaginary part of $t$ as
is obvious in eq.(D.1).} 
 the $\theta$ angle as
$\theta \r \theta +2 \pi$. This shift is an exact symmetry for a GLSM
on a compact world sheet without a boundary (because ${1\over 2 \pi}\int v$ 
is integrally quantized in such a theory). However its action on boundary 
conditions is nontrivial, (recall flux is not quantized for a 2d gauge
theory on the disk). In particular in the UV of the GLSM RG flow,
fractional instantons (configurations in which the phase of the vev of
$\ph_n$ winds once on the boundary of the disk) violate this shift
symmetry, and distinguish\foot{In fact the one
point function of the bulk ${\bf C/Z}_n$ tachyon, on the disk, with boundary 
conditions corresponding to the fractional brane $|i\rangle$ is given
by (see \DoMo \DiGo \Bi)
\eqn\bo{c_{m}={\cal N} \ e^{-2\pi i m/n}.} 
This one point function
corresponds to the GLSM partition function in the presence of a single 
fractional instanton, which is clearly proportional to  
$e^{-i\theta/n}$, consistent with the assignment $\theta = 2\pi m$ for
the brane $|m\rangle$. }  
the $n$ different D2-branes.  

On shifting $\theta$ as described above the boundary condition continues 
to be ${\cal N}=2$ supersymmetric. In fact shifts in $\theta$ are the only 
${\cal N}=2$ supersymmetric boundary conditions generated by 
marginal or relevant 
boundary operators (spacetime gauge fields). As the $\theta$ angle is not 
renormalized under RG flow, we conclude that the fractional branes
$|m\rangle$ flow in the IR to the GLSM with Neumann boundary conditions and 
$\theta= 2 \pi m$. However the fractional instantons that allowed one to
distinguish between $|m\rangle$ in the UV are absent in the IR (because the 
$\ph_1$ rather than $\ph_n$ has a vev in the IR). Consequently $\theta$ is 
effectively periodic with periodicity $2 \pi$ in the IR, and all the $n$ 
branes $|m \rangle$ evolve into the same bulk flat space D2-brane.

In order to develop intuition for these flows, we now study the 
evolution of the spacetime gauge fields on the $m^{th}$ (fractional)
D2-brane in the `classical' sigma model approximation. As we have 
explained in Section 3, the world-sheet gauge field $v_0$ is easily 
computed as a function of scale in this approximation. However $v_0$ is 
simply related\foot{This is shown by the identification 
of the boundary coupling 
\eqn\binti{m \int dx^0 ({\sigma + {\bar \sigma} \over 2} - v_0),}
with the usual sigma model gauge coupling $\int A^\mu(X)\de_0 X^\mu$ setting
$\sigma=0$ in the Higgs branch. This gauge coupling comes from the 
shift of the parameter 
$\theta$ in the term $\f{\theta}{2\pi}\int v$. The $\sigma$ dependent term
is also induced to preserve ${\cal N}=2$ supersymmetry (see \Ho). 
} to the spacetime gauge field; 
 $m v_{0}=A_\mu(X)\de_{0}X^\mu$ 
\HoIqVa \Ho. Using these formulas we find that,
for $r>0$ (in the UV region\foot{
Under the RG-flow the FI parameter $r$ changes following (B.2) 
as reviewed in the Appendix B. The UV limit
corresponds to $r=\infty$, while IR limit to $r=-\infty$.}) 
the flux is delta functionally localized,  
\eqn\gfluxuv{F=-m \delta(\rho)\ d\rho\we d\theta.}
On the other hand when $r<0$ 
\eqn\gflux{F=-\f{2 m  n|r| \rho}{\left(n(n+1)\rho^2+|r|\right)^2}\ 
d\rho\we d\theta.}
Thus, as the RG-flow proceeds, the flux \gflux\ (whose integral is
conserved) expands out with the bubble of flat space. 
 In the IR limit the flux is completely
diluted and the $n$ D2-branes are indistinguishable. 

This property of boundary RG-flow can also be seen in 
the sigma model with boundary coupling 
$\int_{\de\Sigma} A_{\mu}(X)\de X^\mu$. The one loop RG-flow
equation \CallanBC\ leads to (in the weak curvature approximation)
\eqn\rga{\f{\de A_\mu}{\de \lambda}=\nabla^\nu F_{\nu\mu}-\xi^\nu 
F_{\mu\nu}+\nabla^\nu \phi\  F_{\mu\nu},} 
where $\lambda$ is the logarithm of length scale (proportional to 
$|r|$ in \gflux) 
under the flow and 
$\xi^\nu$ represents the freedom of coordinate transformations. 
After we substitute the 
results\foot{
They are given by the metric $(ds)^2=\lambda(f(s)^2ds^2+s^2 d\theta^2)$
and the parameter $\xi=\f{1}{2}sf(s)ds$ with zero dilaton \GuHeMiSc.} 
of bulk RG-flow in \GuHeMiSc, we obtain the following 
equation assuming that only $A_\theta$ is nonzero,
\eqn\rga{\f{\de A_\theta}{\de \lambda}=\f{s}{f(s)\lambda}\ \de_{s}
\Bigl(\f{1}{f(s)s}\de_{s}A_\theta+\f{1}{2}A_\theta\Bigr),}
where we rescaled the coordinate as $s=\rho/\sqrt{\lambda}$ and defined
the function $f(s)$ by $f(s)=(1+W(\f{1}{n}-1-s^2/2))^{-1}$
($W$ is the inverse function of $xe^x$). By setting the right-hand side of 
\rga\ to zero, we find a solution (up to constant shift)
\eqn\sola{
A_{\theta}=A_{\theta}(0)\cdot e^{-\int^{s}_0\f{s'f(s')}{2}ds'}.}
After transforming into the previous radial coordinate $\rho$
we get the gauge flux
\eqn\solf{F_{\rho\theta}=-\f{A_{\theta}(0)\cdot f(\rho/\sqrt{\lambda})
\cdot \rho/\sqrt{\lambda}}{2\sqrt{\lambda}}
\cdot e^{-\int^{\rho/\sqrt{\lambda}}_0\f{s'f(s')}{2}ds'}.}
Since the function $f$ satisfies the bound $1/n<f<1$, it is obvious that
the flux $F_{\rho\theta}$ value takes its the maximum at 
$\rho\sim O(\sqrt{\lambda})$ as in \gflux\ 
and the asymptotic behavior ($r\to \infty$)
is similar to the solution to diffusion equation as
$F_{\rho\theta}\sim e^{-\f{f\rho^2}{4\lambda}}$. 
Thus it is completely smeared
in the IR limit $\lambda\to \infty$ and we again confirmed the 
diffusion property in a different picture.

It is instructive to analyze how the spectrum of open strings between 
the fractional branes $|0 \rangle$ and $|m \rangle$ evolves as a function
of RG scale. In the UV the relative (spacetime) field strength between these
two branes is given by \gfluxuv\ (or equally, a Wilson line 
\eqn\wilsonl{\int _{0}^{{2\pi \over n}} A_{\theta} d \theta= -{2 \pi m
\over n},}
in the `angle' direction). 
Consequently, open strings stretching between these branes obey twisted 
boundary conditions. For instance the (open string) tachyon field obeys
\eqn\pro{
T(z)=e^{-2\pi im /n}\ T(gz),}
where the operator $g$ acts on the coordinate $z$ of ${\bf C}/{\bf Z}_{n}$ 
as $g:z\to e^{2\pi i/n}z$ in agreement with \bcf. As the RG flow proceeds
towards the IR, however, the flux enclosed within a circle of any finite 
spacetime radius (and hence the effective Wilson line at that radius) 
goes to zero (see \gflux). In other words the world-volume fields obey 
twisted boundary condition like \pro\ for $|z|>> \sq{|r|}$, but satisfy 
the untwisted one inside the bubble of flat space $|z|<< \sq{|r|}$.

\vskip .2in

{\bf D0-branes}

We will now study the evolution of D0-brane boundary conditions under 
the RG flow. Unfortunately it appears difficult to find simple boundary
conditions that both preserve ${\cal N}=2$ and also reduce in the UV to the
D0-brane; however, we can circumvent this problem with a trick. 
As we will explain below, it is straightforward to construct 
${\cal N}=2$ GLSM 
boundary conditions that reduce in the UV to the $D2-{\bar D}2$ system.
These boundary conditions may then be deformed (while preserving 
${\cal N}=2$
supersymmetry) by turning on a vortex for the open string tachyon. The
resulting boundary condition may be interpreted as a single D0-brane (the 
$D2-{\bar D}2$ brane have disappeared via open string tachyon 
condensation \SenSM). 
Non renormalization of the boundary super potential 
(guaranteed by ${\cal N}=2$ supersymmetry) allows us to follow the evolution
of the boundary conditions as a function of RG scale, and determine the
fate of fractional D0-branes under tachyon condensation. 

To be more concrete, let us consider the $D2 \bar{D}2$ system 
in ${\bf C/Z}_n$ 
where the $D2$ brane is of type $|0\rangle$ and the ${\bar D} 2$
brane is of type $|m \rangle$ from the viewpoint of GLSM.
As we have seen before, the action of ${\bf Z}_{n}$ quantum symmetry
(or equally the shift of $\theta$ by $2\pi$) changes the types
of 2-branes in UV limit. Since the shift of $\theta$ by $2\pi n$ is the 
symmetry of GLSM with boundary in the orbifold limit, we have the 
periodicity $|m+n \rangle=|m \rangle$ at the beginning of the RG-flow.

In order to construct boundary conditions that represent the $D2-{\bar D}2$
system, we must introduce Chan Paton factors. This is conveniently achieved
by introducing a new complex fermionic boundary variable $\eta$. It will 
turn out (see below) that $\eta$ and its conjugate ${\bar \eta}$ obey 
the conjugation relations $\{\eta,\bar{\eta}\}=1$; consequently
quantization leads to a two state system. Operators on this Hilbert space
are $2 \times 2$ matrices; as expected for the $D2 \bar{D}2$ system. 

Then the $D2 \bar{D}2$ system in ${\bf C/Z}_n$ 
 is described by the GLSM with boundary 
conditions \bcdtt, together with gauge invariant 
boundary interactions (this can also 
be seen as the 
boundary string field theory \WittenBSFT \BSFT\ 
for brane-antibrane system \DD)
\eqn\BSFT{
S_{\de\Sigma}=\int_{\de\Sigma} d\tau d\theta d\bar{\theta} \
\bar{\Gamma}e^{-mV}\Gamma
+\int_{\de\Sigma} d\tau d\theta\ \Gamma T(\Phi)+(h.c.),}
where $\Gamma$ is a charge $m$ boundary fermionic chiral superfield\foot{
In the UV limit this charge means that the ${\bf Z}_n$ orbifold 
action acts as
$\Gamma\to\Gamma\ e^{2\pi i/m}$. Thus the tachyon followed the projection
$T\to T\ e^{-2\pi i/m}$. This fact agrees with \pro .}
whose 
lowest  component is $\eta$ 
\eqn\bdyf{\Gamma=\eta+\theta F-i\theta\bar{\theta}\partial_{\tau}\eta ,}
and $V$ is the GLSM vector superfield whose lowest component is given by
$(v_0-\f{\sigma+\bar{\sigma}}{2})\theta {\bar \theta}$. 
In components, the first term in \BSFT\ is simply 
\eqn\detint{
-m \int {\bar \eta} \eta (v_0 -{\sigma + {\bar \sigma} \over 2})
\sim -m \int \f{\sigma_3}{2} (v_0 -{\sigma + {\bar \sigma} \over 2}),}
the analogue of \binti\ on a single D2-brane. Note, in particular, that 
$v_0=\de_{\tau}X^\mu A_{\mu}(X)$\foot{The scalar field $\sigma$, which is 
in the same ${\cal N}=2$ multiplet as $v_\mu$, is taken to be zero
since we are interested in the Higgs branch. For the analysis
of the Coulomb branch in similar GLSMs see \MaMo.} 
where $A_{\mu}$ is the relative gauge field in the $D2 \bar{D}2$ system.
The second term in \BSFT\ represents the boundary deformations 
corresponding to the open-string tachyon field $T(\Phi)$, where $T$ is a
complex scalar.

It turns out that the GLSM with boundary conditions \bcdtt\ and boundary 
interactions \BSFT\ preserves ${\cal N}=2$ supersymmetry provided 
that the tachyon 
$T$ is a holomorphic function of $\Phi$. The simplest solution to this 
condition is $T=0$; this represents a $D2-{\bar D}2$ system in
the UV, and flows to a usual 
$D2-{\bar D}2$ system in the IR. In particular, the 
open string tachyon, which is protected against quantum corrections 
owing to a non-renormalization theorem, remains zero throughout the flow. 

{~~~}

{\it Flows from bulk 0-branes into bulk 0-branes}

We now consider boundary conditions with the open string tachyon turned on. 
We first consider the case $m=0$ (i.e. 
the 2-brane and anti 2-brane are of the same variety).
In this situation $\Gamma$ is uncharged; consequently gauge invariance of 
\BSFT\ requires that $T(\Phi)$ should be gauge invariant (uncharged). 
The simplest example is given as follows
\eqn\stak{
T(\Phi_{-n},\Phi_{1})=T_{1}\Phi_{-n}(\Phi_{1})^n,}
where the coefficient $T_{1}$ corresponds to the strength of the
open-string tachyon field. In the UV limit the field $\Phi_{-n}$ has a large
expectation value and can be replaced by $\sq{r}/n$, while $\Phi_{1}$ can be
regarded as a coordinate of the orbifold ${\bf C}/{\bf Z}_n$. Thus the 
tachyon field is expressed\foot{Note that this 
field configuration satisfies the projection \pro\ 
setting $m=0$.}
\eqn\takw{
T(z)=\f{T_{1}\sq{r}}{n}z^n.}

 In the deep UV $T(z) = k z^n$ where $k$ diverges; 
and so represents a bulk D0-brane \ta\ at the origin of the orbifold. 
(The tachyon field \stak\ has winding number $n$ and so would have
represented $n$ D0-branes in flat space. A bulk 0-brane on ${\bf C/Z}_n$ is 
precisely a collection of $n$ covering space 0-branes modded out by the 
projection, explaining our identification).

On the other hand, in the IR limit, the field $\phi_{1}$ has 
an expectation value  $\phi_{1} \approx \sqrt{|r|} $ 
, and another field 
$\phi_n$ is identified with $z$ the coordinate in flat space. 
The tachyon field configuration \stak\   
\eqn\takz{
T(z)=T_{1}\sq{|r|} z ,}
has unit winding number and so may be identified with a D0-brane in the 
flat space. Thus we conclude that a bulk D0-brane evolves into an
ordinary D0-brane the closed string tachyon condensation, consistent 
with the results of subsection 5.1. 

This analysis is easily extended to more complicated 
tachyon configurations.
By following the RG flow in the presence of the polynomial 
valued tachyon field
\eqn\takcom{
T(\Phi_{-n},\Phi_{1})=\sum_{l=0}T_{l}(\Phi_{-n}(\Phi_{1})^n)^l,}
we conclude that a configuration of $n$ separated bulk D0-branes 
(located at the zeros of $\sum_{l=0}T_{l}w^l=0$)  evolve into $n$ 
separated D0 branes in flat space.  

{~~~}

{\it Flows from bulk 0-branes into nothing}

We can also choose the following tachyon field for the same kind of
$D2-\bar{D}2$ system setting $m=n$ (remember the periodicity 
$| m+n \lb=| m \lb$ in the UV limit)
\eqn\stakbn{
T(\Phi_{-n},\Phi_{1})=T_{1}(\Phi_{1})^n.}
Even though the starting point is the same as before (a bulk D0-brane),
the final product is different. Since the tachyon field takes a 
large constant value and is completely condensed in the end, 
the bulk D0-brane disappears in this process. Note also that this only 
happens near the origin ($z=0$) because
it is 
impossible to shift the location of the D0-brane away from the origin
as in \takcom. These results are completely consistent with those 
obtained from the geometrical consideration in section 5.1.

{~~~}

{\it Flows from fractional 0-branes into nothing}

We now turn to the study of tachyon condensation on the more general 
system consisting of a D2-brane $| m \lb_{D2}$ and an anti D2-brane
$|\bar{0}\lb_{D2}$. In this situation $\Gamma$ has charge $-m$, and 
charge conservation demands that $T(\ph)$ have charge $m$. 
The simplest example of such a tachyon field is
\eqn\takfr{
T(\Phi_{-n},\Phi_{1})=T_{1}(\Phi_{1})^{m}.}
In the UV the tachyon field $T(z)\sim z^{m}$ may be interpreted as 
representing $|m  \lb_{D0},|m-1\lb_{D0},\ddd,|1\lb_{D0}$
(this follows from an analysis of the RR charges of this configuration, 
see \ta). On the other hand, in the IR limit $\phi_1 \approx \sqrt{|r|}$ and
the tachyon field is simply constant (and this constant value increases
without bound as we flow to the deep IR). Thus the open string tachyon 
is completely condensed; all open string degrees of freedom vanish in this
limit; consequently the fractional 0-brane has simply disappeared.
This matches the results of section 5.1 again.

{~~~}

{~~~}

{\it 0-brane production from nothing}

Finally, let us consider the case of $m=-n$ 
(the same kind of D2 and antiD2). 
Then an allowed tachyon field is
\eqn\takflux{
T(\Phi_{-n},\Phi_{1})=T_{1}\Phi_{-n},}
which shows the initial nothing state will become a D0-brane
in the flat space after the closed string tachyon condensation. 
Once again, this also matches with the previous geometrical argument
and corresponds to the cycle ${\bf S}^1_{B}$.

\subsec{GLSM Analysis of Twisted Circle Boundary Flows}

As we have described in section 3, the decay of twisted circle theories 
via closed string tachyon condensation could easily be understood utilizing 
a GLSM. The low energy limit of the GLSM is a CFT with a Liouville factor;
the Liouville direction may (roughly) be thought of as a scale or Euclidean
time. In this subsection we will study the same GLSM on a world sheet with 
a boundary, imposing boundary conditions that preserve 
${\cal N}=2$ B-type world sheet
supersymmetry. The boundary conditions we study will always be Neumann in 
the Liouville ($P_1$) direction (this is required for 
${\cal N}=2$ world sheet 
supersymmetry, \SuYa \EgSu); and in most examples we study the boundary 
interaction will also be independent of $P_1$. As in the previous
subsection, ${\cal N}=2$ supersymmetry guarantees these 
features persist in the 
low energy CFT; it is then a simple matter to follow the boundary
conditions, and hence trace the D-branes from $P_1 \gg 0$ (twisted circle) 
to $P_1 \ll 0$ (supersymmetric circle). As the arguments of this
subsection closely mimic those of the previous subsection, we will be very 
brief. 

The decay of a twisted circle D3-brane is described by a D4-brane in the 
GLSM of section 3. The twisted circle D3-brane with zero Wilson line is 
described by Neumann boundary conditions on $\ph_1$, $\ph_n$, 
$P_1$ and $P_2$, together with the conditions $v_1=v_{01}=0$ on the
boundary (these conditions preserve ${\cal N}=2$ B-type world sheet
supersymmetry). Turning on a Wilson line corresponds to adding the boundary 
interaction \binti. The UV Wilson line ${ 2 \pi m \over n}$ 
($m$ in \binti\ varies continuously between $0$ and $n$) evolves 
into\foot{Taking the zero radius limit (see 
subsection 4.2) we conclude that all ${\bf C/ Z}_n$ fractional
2-branes map into the bulk flat space 2-brane, in agreement with the 
previous subsection.} the 
IR Wilson line ${ 2 \pi m}$. 

As in the previous subsection, it is possible to construct the twisted
circle fractional $D1$ brane ($D2$ brane on including 
the Liouville direction) via $D3 {\bar D}3$ ($D4 {\bar D}4$ on including
the Liouville direction) tachyon condensation. $P_1$ independent boundary 
deformations of the $D4 {\bar D}4$ are identical to \stak\ for a bulk 
D-brane and \takfr\ for a fractional D-brane. The arguments and conclusions
of the previous subsection carry through almost unmodified, once again
confirming the geometrical reasoning of subsection 5.1. In particular, we
find that the bulk D1-brane in the twisted circle either disappears or 
evolves into an ordinary 
D1-brane in flat space after the closed string tachyon condensation. On the 
other hand, the fractional D1-brane disappears from the spectrum.

Finally we should point out that it is also possible to  construct large
classes of ${\cal N}=2$ boundary interactions which depend on $P$. 
For example, we can take 
\eqn\ptab{
T(P,\Phi_{-n},\Phi_{1})=T_1(\Phi_{1})^{m+q}e^{-qP}.}
These are expected to lead to other boundary states of the GLSM 
after the boundary RG-flow\foot{Here we should consider relevant
perturbations $q<\f{k}{4(n-1)}$ or marginal one $q=\f{k}{4(n-1)}$.
}. We will not investigate the
relevance of these D-branes to our closed string tachyon condensation.

\newsec{Evolution of D-branes under More General RG Flows}

Before we conclude this paper, we would like to apply our method of 
analyzing D-brane spectrum to more general RG flows. In particular,
let us consider\foot{We are very grateful to A. Adams and R. Gopakumar
for suggesting an investigation of this case to us.}
 the RG Flows from 
${(\bf C\times S^1)/Z}_{n}$ to ${\bf( C\times S^1)/Z}_{m}\ \ (n>m)$. We 
further
assume that $n$ and $m$ are coprime. Then we can construct gauged linear
sigma models which describe these RG-flows in the same way as before. 
This is defined by just replacing the
chiral superfields $\Phi_{1}$ in \ft\ with the charge $m$ superfields
$\Phi_m$ in the model discussed in section 3.1. If we take the small radius
limit $k\to 0$, then it reduces to the RG-flow from 
${\bf C/Z}_{n}$ to ${\bf C/Z}_{m}$, which is also realized as a gauged 
linear sigma model in the same way \Va.

Now let us consider the evolution of D-branes under 
these RG-flows\foot{Even though we use the terminology of 
D-branes in the twisted circle theory, 
we can also get the equivalent result for the D-branes in the
orbifold.}. For simplicity, we consider the explicit example of 
$(n,m)=(5,3)$, 
though its generalization is obviously straightforward. Again we can apply 
two different methods. 

First, we analyze the geometrical picture 
as in section 5.1. We can compute the classical metric as in Appendix C 
and indeed get smooth four geometry. In the UV limit it 
approaches the metric
of ${(\bf C\times S^1)/Z}_{n}$ given by
\twim, while in the IR limit it becomes the metric of 
${(\bf C\times S^1)/Z}_{m}$
\eqn\cnm{ds^2=d\rho'^2+\f{n^2}{m^2}\rho'^2d\theta^2
+\f{k}{2}\Bigl(\f{d\theta}{m}-dP_2\Bigr)^2.} Note that the angle 
$\theta(\equiv \theta_m)$ has the periodicity $2\pi$ and also we have
the ${\bf Z}_{n}$ identification 
$(\theta,P_2)\sim (\theta+\f{2\pi m}{n},P_2+\f{2\pi}{n})$.

Consider a fractional D1-brane wrapped on the twisted circle 
${\bf S}^1_A$ in the UV 
defined in the same way as in section 3.3 (see Fig.5).
After the RG flow, it again disappears as we can see from \cnm\ explicitly.
However, a bound state of two fractional D1-branes wraps a cycle 
equivalent to ${\bf S}^{1}_{D}$ (equivalence upto the shift of
the UV trivial cycle ${\bf S}^{1}_{B}$). It follows from \cnm\ that 
the cycle ${\bf S}^{1}_{D}$ degenerates at $\rho'=0$, 
but then evolves into the twisted circle of ${(\bf C\times S^1)/Z}_{m}$ upon
subsequent evolution. Thus we predict that even though 
a single fractional D1-brane disappears under tachyon condensation, 
a bound state of two such branes can either evolves into
a fractional D1-brane or disappear upon closed tachyon condensation. 
On the other hand, if we consider a bulk D1-brane wrapped on 
${\bf S}^{1}_{C}$, it will simply evolve into a bulk D1-brane in the IR 
twisted circle as in the previous case of $m=1$.

\fig{Various cycles in the twisted circle ${(\bf C\times S^1)/Z}_{5}.$}
{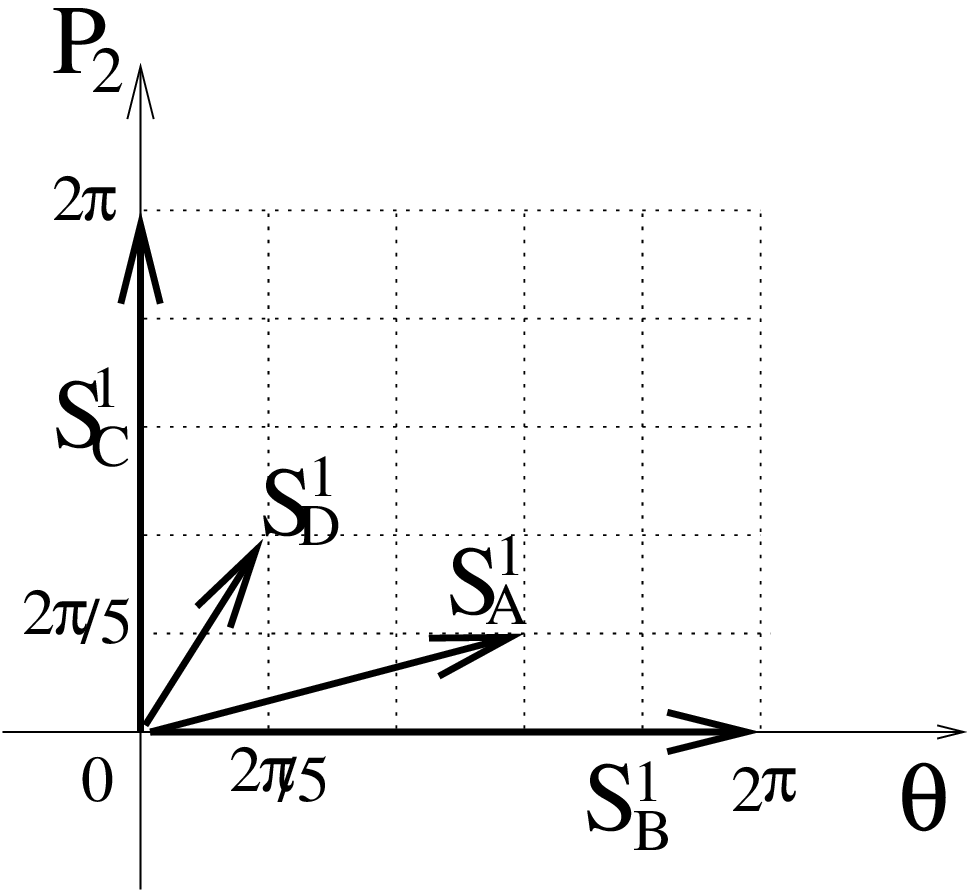}{2.5truein}

Next let us understand the above result by employing 
the analysis of ${\cal N}=2$ supersymmetric 
boundary interaction. The tachyon configurations
\eqn\tnmdo{T=\ T_1 \Phi_3,\ \ \ T_1 (\Phi_3)^2,\ \ \ T_1 (\Phi_3)^2 \Phi_{-5},
\ \ \ T_1 (\Phi_3)^5 (\Phi_{-5})^3, \ \ \ T_1 (\Phi_3)^5, \ \ \
T_1 (\Phi_{-5})^3,  }
respectively represent
\item{1.} The disappearance of a single fractional D0-brane.
\item{2.} The disappearance of the bound state of two fractional D0-branes.
\item{3.} The evolution of a bound state of 2 fractional D0-branes into 
          a single fractional D0-brane.
\item{4.} The evolution of a bulk D0-brane into a bulk D0-brane.
\item{5.} The disappearance of a bulk D0-brane. 
\item{6.} The creation of a bulk D0-brane out of nothing.

Note the striking match between geometrically inspired conjectures and 
our explicit constructions of boundary RG flows. Indeed, the tachyon fields
\tnmdo\ explicitly 
encode the information of the lattice vectors (see Fig.5) via
the `translation rule'
$\Phi_m\lr {\bf S}^{1}_{A}$ and $\Phi_{-n}\lr -{\bf S}^{1}_{B}$. 
This obviously allows us to see the results of D-brane evolutions under the
general $(n,m)$ RG flow\foot{The evolution of D3-branes can also be 
discussed in the GLSM picture. A D3-brane is parametrized by the 
Wilson line $\f{2\pi l}{n}\ \ (0\leq l< n)$ in the UV limit. 
After the RG flow, the periodicity becomes small as $0\leq l< m$ by the 
radius change.}
.

\newsec{Discussion}

In this paper we have studied the evolution of D-branes under closed string
tachyon condensation in non-supersymmetric orbifold and twisted 
circle backgrounds. Our analysis employed two complementary methods. 
First, we used the gauge linear model description of the twisted
circle tachyon condensation process to derive a smooth four dimensional
geometry that approximately describes the evolution of the twisted circle
in an energy scale (roughly Euclidean time) direction. Interpreting 
the branes of the twisted circle theory
as usual type II branes wrapping nontrivial cycles, we traced the evolution
of these cycles as the tachyon condensed. Our second method involved 
the analysis of RG flows in gauged linear sigma models on the disk; 
we used world sheet ${\cal N}=2$ supersymmetry to follow the evolution of 
boundary interactions (sometimes using ideas from boundary string field 
theory) into the IR.  Both methods yielded the same result; 
localized D-branes disappear (or 
evolve other fractional D-branes in general RG flows) as the closed string
tachyon condenses, while D-branes that are not tied to the fixed point 
either disappear or evolve into usual D-branes in flat space. 

Our study suggests several directions for future study. First, twisted
circle tachyon condensation is only approximately described by the four
dimensional
Euclidean geometry studied in this paper. The exact process is captured by 
flowing the sigma model based on this approximate
geometry to the IR. It would be intensely interesting  
to find an exact description of the resultant ${\cal N}=2$ CFT. Indeed 
Hori and Kapustin \HoKa\ were able to solve a related problem in the simpler 
setting of 2d Black hole. 
Such a description would permit detailed understandings 
of several 
aspects of the process of tachyon condensation. 

Secondly, we have only studied tachyon condensation as a function of 
the Liouville direction (roughly Euclidean time). It is certainly important
to understand the Lorentzian analogue of this process (the evolution of this
background as a function of time). It is not clear to us that the
Euclidean and Lorentzian processes are even qualitatively similar; for instance
the geometry on equal `time' ($P_1$) cycles of our smooth 4-geometry 
\metrico\ degenerates at the slice $P_1=0$. It would be very interesting to 
understand this issue better\foot{See \GuHeMiSc\ for a recent discussion on
the connection between world sheet RG flow and time evolution.}.

Our second approach to this problem also highlights several interesting
connections. As we have explained the world sheet RG flow with certain 
boundary conditions (for instance boundary conditions corresponding to
the fractional 0-brane of ${\bf C/Z}_n$) leads to world sheet theory with 
no boundary in the IR. As we have explained in subsection 5.3, there is a
sense in which closed string tachyon condensation triggers open string
tachyon condensation on an initially stable brane. Intuitively, it is not 
difficult to understand why this happens. For instance, 
the fractional 0-brane of ${\bf C/Z}_n$ may be `blown up' into a D1-brane
wrapping a twisted circle. This circle (non contractible in the UV)
degenerates in the process of the closed string tachyon condensation. The
degeneration of this cycle allows the $D1$ brane circular loop to self
annihilate via open string tachyon condensation (analogous to 
$D1 \bar{D}1$ annihilation). As is remarkable even in the general RG flow
analysis in section 6, we can obtain the same conclusions 
from our two different viewpoints. This might be mathematically 
regarded as a novel correspondence
(analogous to McKay correspondence\foot{See also \MaMo \YHE\ for 
similar discussions in four dimensional unstable orbifolds.})
between geometrical structure and 
algebraic one under `time' evolution.

Another interesting direction of investigation concerns the limit 
$n \to \infty$ of the models studied in this paper. As we have explained in 
Section 2, in this limit the twisted circle reduces to an orbifold of 
type 0 theory. If the study of tachyon condensation commutes with $n \to
\infty$, the results of this paper determine the fate of several 
type 0 D-branes under the process of closed string tachyon condensation. As 
the type 0 tachyon is delocalized, type 0 tachyon condensation is 
particularly interesting, and deserves further study.

Finally, in this paper we have considered configurations consisting of a
fixed number of D-branes, in the classical limit $g_{string} \to 0$. In
this limit D-branes probe (and react to) the closed string geometry, but do
not back react on it. New interesting phenomena may occur in other limits;
for instance if the number of branes is taken to infinity. We leave such
investigations to future work.

\vskip 0.4in

\centerline{\bf Acknowledgments}

\vskip .1in

We would like to thank J. David, T. Eguchi, M. Gutperle, 
Y.Hikida, A. Iqbal, T. Kawano, Y. Michishita, A. Sen, E. Silverstein, 
A. Strominger,  H. Takayanagi, D. Tong, N. Toumbas, C. Vafa,  S. Yamaguchi, 
P. Yi and especially  A. Adams, R. Gopakumar, M. Headrick and K. Hori for 
several extremely useful discussions. We would particularly like to thank 
M. Headrick for disabusing us of a potentialy embarrassing misconception. The
work of SM was supported  in part by DOE grant DE-FG03-91ER40654
and a Harvard Junior Fellowship. The work of TT was
supported in part by DOE grant DE-FG03-91ER40654.

\appendix{A}{Type 0 theory and Interpolating Orbifolds}

Type 0A/B theory may be regarded as an orbifold by $(-1)^{F_s}$ ($F_s$ is
the spacetime fermion number) of  type IIA/B theory. This is most easily 
seen at the level of the torus partition
function. The projection by $(-1)^{F_s}$ retains only states in the
NS-NS or RR sectors of the theory, i.e. states with fermion 
boundary conditions
$(-,-)$ or $(+,+)$ on the $\sigma$ cycle of the torus. It now remains to 
add in the twisted sectors; as the net result of this process is symmetric 
(i.e. modular invariant) 
between the two cycles of the torus, the final partition function thus 
contains only $(+,+)$ and $(-,-)$ along the $\tau$  cycle as well and 
the resulting partition function is that of type 0A/B.

Recall, that type 0A and type 0B theory are T-dual to each other; 
under T-duality $\psi_9 \r - \psi_9$, and so $(-1)^{F_l+F_r} \r
-(-1)^{F_l+F_r}$ in the Ramond sector (this is most clearly seen on 
expressing $(-1)^{F_l+F_r}$ in terms of bosonized fermions),
interchanging type 0A and type 0B.

We now move onto the consideration of a slightly more involved orbifold
of type IIA/B theory; the orbifold by $(-1)^{F_s} \times \sigma$ where
$\sigma$ represents a shift by $2 \pi R$ in the $z$ direction \CoGu.
As we now review, in the $R \to 0$ limit this so called `interpolating 
orbifold' is equivalent to type 0B theory on ${\bf R}^{10}$; it
interpolates  between IIA theory on 
${\bf R}^{10}$ (as $R \to \infty$ ) and
0B theory on ${\bf R}^{10}$ (as $R \to 0$). It is clear that the projection
by $(-1)^{F_s} \times \sigma$ retains only fermions of half odd momentum
and bosons of integral momentum around the $z$ circle. At the level of the 
torus partition function, the sum over windings of the $\tau$ cycle over 
the spacetime circle (parameterized by the winding number $w_{\tau}$) is
weighted by the additional factor $(-1)^{w_t}$ in the $(-,+)$ or 
$(+,-)$ sectors. Once again, the symmetry of the twisting procedure 
between the $\sigma$ and the $\tau$ cycles demands that the summation over 
windings in the $\sigma$ direction be weighted by $(-1)^{w_s}$ when the
fermions have $(-,+)$ or $(+,-)$ boundary conditions in the $\tau$
direction. This reverses the GSO projection, setting 
$(-1)^{F_L}=(-1)^{F_R}=1$, (instead of the usual $(-1)^{F_L}=(-1)^{F_R}=-1$)
in sectors of odd winding. 

In summary, this orbifold retains 
only fermions of half odd momentum and bosons of integral momentum. 
Furthermore, the GSO projection is reversed in sectors of odd winding.
It is now clear that the orbifold of IIA theory by $(-1)^{F_s} \times \sigma$
reduces, as $R \to 0$, to 0B theory in 
noncompact ${\bf R}^{10}$. In this limit all the states of non-zero 
momentum around the circle are lifted out of the spectrum and this  
includes all the fermions (since by the antiperiodic boundary 
conditions they have half-integral momenta). In the sector with 
zero-momentum along the circle,  
the only effect of the boundary conditions is to reverse the GSO projection 
for states of odd winding number $w$.
Thus when $w=2w'$ ($w$ is the winding number) the spectrum consists of the
(NS+,NS+) and (R+,R-) sectors, and for $w=2w'+1$ the (NS-,NS-)
and (R-,R+) sectors. This is approximately the spectrum in the
zero-momentum, winding number $w'$ sector of type 0A theory on a circle of
radius $2R$. This equality becomes exact in the limit $R \to 0$, where it is
more appropriate to describe the theory as 0B in infinite flat space.

We now describe the T-dual of the interpolating orbifold. T-duality
is easiest to understand in terms of states rather than partition functions.
As usual, T-duality
interchanges momentum and winding. The T-dual theory is most usefully 
thought of as a theory with radius ${\apm \over 2R}$. Sectors of odd winding
are all fermions, sectors of even winding are all bosons. In addition, 
states of integral momentum have the IIB/A GSO projection, while states of 
half odd momentum have the reversed GSO projection. In the limit that
$R \to 0$, this theory is approximately type 0B/A theory on a circle of
radius ${\apm \over 2R}$. At finite $R$ the model is an orbifold of
type 0B/A theory by $(-1)^{F_L} \times \sigma'$ where $\sigma'$ is
a shift by  ${2 \pi \apm \over 2R}$ in the $z$ direction. The projection
associated with this orbifold ensures that states with the usual (IIB)
GSO projection have integral momentum; states with the opposite 
GSO projection
have half odd momentum. Twisted sectors then produce the rest of the
spectrum, as may be seen, as usual, by demanding symmetry of the partition
function under $\sigma \leftrightarrow \tau$.

\appendix{B}{Decay of ${\bf C}/{\bf Z}_n$ }

\subsec{Review of Tachyon Condensation in ${ \bf C/Z}_n$}

Recall that type II theory on ${\bf C}/{\bf Z}_n$ has tachyons 
in its twisted
sectors. Adams Polchinski and Silverstein (APS) argued that the tachyon
condensation process in this background nucleates a bubble of flat space at
the tip of the cone \AdPoSi; the bubble then grows without bound. 

Let us consider the actual process of tachyon condensation in more detail. 
Let the initial state of the system be given by the ${\bf C}/{\bf Z}_n$
background perturbed by $\int d^2z ~ \ep~ T$ (in $0-$picture), 
where $T$ is the most relevant 
twisted sector operator. The subsequent evolution of the system under 
worldsheet RG flow has two distinct stages. The first stage (the stage
of linear growth) is characterized by the condition $\ep \ll 1$. In this 
stage $\ep$ simply grows with RG scale at a rate determined by the 
$2(1-\delta)$, where $\delta$ is the holomorphic scaling dimension of $T$; 
no other operators are turned on. Eventually $\ep$ grows to unit order, and 
begins to mix in a nonlinear fashion with all other operators. At this 
point the spacetime metric (hence target space geometry) begins to evolve
with RG scale, and we enter the second stage of the RG flow. 
According to the APS conjecture, in this stage a large bubble of flat two 
dimensional space is nucleated about the tip of (what was) the ${ \bf
C/Z}_n$ cone, and this bubble then grows without bound. 
Once the size of the bubble is large in string unit, the further evolution 
of the system is accurately described by the 1-loop RG equations of the 
worldsheet nonlinear sigma model. 

In order to substantiate this picture, APS first computed the metric on the 
moduli space of D0-branes in the ${\bf C}/{\bf Z}_n$ background perturbed by 
$\int d^2z \ep T$, where $\ep \ll 1$. They found that the 0-brane moduli 
space is ${\bf C}/{\bf Z}_n$ with the singularity smoothened out at 
length scale $\ep \sqrt{\apm}$. This demonstrates that D-branes already 
see the first stage of the RG flow in terms of a growing 
bubble of flat space. Once the size of this bubble becomes large in string
units, a fundamental string must see the same geometry as the D0-brane;
consequently this computation provides evidence for the conjecture
described above. In order to further substantiate their picture, 
APS went on to find an approximate solution
\foot{An exact solution with the same property was
later found in \GuHeMiSc.} to the 1-loop worldsheet 
RG equations on the string
worldsheet (valid once the flat space bubble is large in string units) 
corresponding to this bubble steadily expanding in a sea of 
${ \bf C/Z}_n$. 

In summary, according to the APS picture, D-branes see the tachyon
condensation\foot{Most recent studies of the subject utilize worldsheet RG 
flow (or Liouville flow) to study tachyon condensation. A recent discussion 
of the connection between world sheet RG flow and the physically more
interesting time evolution, may be found in \GuHeMiSc.} process as an
always expanding bubble of flat space. Closed strings see a two stage 
process. In the first stage, the tachyon grows but all other operators 
including the metric are unaffected, and (geometrically speaking) nothing 
happens. Eventually, however, (when the coefficient of the tachyon is 
of order unity, which is also when the D-brane bubble becomes string scale)
the tachyon backreacts onto the geometry, producing a string scale bubble
of flat space, which then proceeds to grow in a diffusive fashion. 

Following up on the work of APS, Vafa \Va\ presented an exact construction 
of the RG flow on the string worldsheet by examining the low energy
behavior of a particular gauged linear sigma model. We will review 
this construction below.

\subsec{Tachyon Condensation using the Gauged Linear 
Sigma Model Construction}

In this subsection we review Vafa's construction of the RG flow describing
${\bf C/Z}_n$ decay utilizing a two dimensional ${\cal N}=(2,2)$ $U(1)$ 
gauge theory (so called gauged linear sigma model (GLSM) \WittenYC\ ). 

Consider a gauge linear sigma model with action 
\eqn\act{S={1\over 2 \pi}
\int d^2\s  \left[d^4\t \left({ \bar \Ph_1} e^V \Ph_1
+{ \bar \Ph_{-n}} e^{-nV} \Ph_{-n}
-{1 \over 2 e^2} |\Sigma|^2\right)+{\rm{Re}}\int d^2\ti{\t}\ t\Sigma
\right],}
where the two charged chiral superfields $\Ph_1$ and
$\Ph_{-n}$ transform under $U(1)$ gauge transformations as in \ft, 
$V$ represents a vector multiplet and its lowest 
component is a gauge field $v_{\mu}$ (its 
field strength is defined to be $v_{01}$); the twisted chiral superfield
$\Sigma$ is its superfield-strength.
The complex valued constant\foot{Actually $r$ is renormalized and runs as
a function of energy scale as we will see later.} 
 $t=r+i\theta$ parameterizes both  the Fayet-Ilipoulos term $r\int d^4\t V$ 
and the theta term $\f{\theta}{2\pi} \int v_{01}$.

The quantum theory generated by \act\ is super renormalizable; 
in fact it is easy to verify that all correlators are rendered
finite after a simple one loop renormalization of the FI term
(see section 3, \WittenYC)
\eqn\fit{r_0(\Lambda_0)=r(\Lambda) +(n-1) \ln 
\Bigl({\Lambda_0 \over \Lambda}\Bigr). }
Here $r_o$ is the bare FI term defined at the cut off scale $\Lambda_0$, 
and $r$ is the finite FI term associated with the finite renormalization 
scale $\Lambda$. 

\subsec{The Sigma Model Limit}

Consider the low energy dynamics of \act. In Wess-Zumino gauge,
the terms involving the auxiliary gauge field $D$ are
\eqn\dterms{{1 \over 2 \pi} \int d^2\s
\left( {D^2 \over 2 e^2}+D(|\ph_1|^2-n|\ph_{-n}|^2 +r) \right),}
where $\ph_1$, $\ph_{-n}$ are the lowest components of
$\Ph_1$, $\Ph_{-n}$ respectively. At energies small compared to $e$
we may restrict attention to fluctuations on
the supersymmetric manifold of zero-energy configurations satisfying
\eqn\ze{-\f{D}{e^2}=|\ph_1|^2-n|\ph_{-n}|^2+r=0.}
The moduli space of classical vacua is parameterized by $\ph_1$ and
$\ph_{-n}$ constrained according to \ze\ modulo gauge transformations,
so that the dynamics is that of a one complex dimensional supersymmetric 
sigma model. 

\subsec{GSO Projection}

We have described above how the GLSM reduces at low energies, to a 
sigma model. We will now describe an exact ${\bf Z}_2$ 
symmetry of the GLSM that 
flows at low energies to $(-1)^{F_L}$ on the sigma model (similar remarks 
apply to $(-1)^{F_R}$). This symmetry may then be used to impose a type II
`GSO projection' on the GLSM.

Classically, the gauge theory
has a discrete chiral symmetry given by $\psi_-  \r - \psi_-$,
$v_\mu \r - v_{\mu}$  $\lambda_+ \r -\lambda_+$
(all other fields are invariant; $v_\mu$ and $\lambda$ are the gauge boson
and the spinor respectively in the gauge multiplet, and $\psi_-$ represent
the negative chirality components of all matter fermions - this a special
example of the left moving R symmetry described, for instance, in
section 2 of \WittenYC.) is a symmetry of
the gauge theory Lagrangian; we call the operator that generates this
symmetry $(-1)^{F_L}$. This change of variables induces a Jacobian
$(-1)^{{n-1 \over 2 \pi} \int v_{01}}$ in the quantum path integral; this
Jacobian is trivial if and only if $n$ is odd.
Thus $(-1)^{F_L}$ is non-anomalous in
the quantum theory if $n$ is odd, as we assume in this paper.

\subsec{The Sigma Model in the UV}

Let us start with the UV limit $r=\infty$.
We can use the
gauge freedom to set $\ph_{-n}$ to be real and positive, and then use
\ze\ to solve for $\ph_{-n}$ as $\ph_{-n} = \sqrt{(|\ph_1|^2 +r)/n}$.
We then plug this solution into the kinetic terms 
\eqn\kintsf{
{1 \over 2 \pi}
\int d^2\s ( -\CD^\m \ph_1 \CD_\m \ph_1 -\CD^\m \ph_{-n} \CD_\m \ph_{-n}),
}
(where $\CD_\m \ph_1$, $\CD_\m \ph_{-n}$ are the usual covariant
derivatives) and classically integrate out\foot{When $r$ is very large the
classical approximation is valid as (according to \ze) $\ph_{-n}$ is also
very large in this limit, and so the gauge boson is very 
massive, by the Higgs
mechanism - recall the gauge theory is free in the UV). } 
the gauge boson $v_\m$ (see
subsection B.7 for the details).
We find that the dynamics in this limit is governed by the flat sigma model
$S={1\over 2 \pi} \int d^2\s \left(
\p_\m \ph_{1} \p^\m { \bar \ph_1} \right)$.
Note, however, that our gauge choice leaves unfixed a
residual ${\bf Z}_n$ group of gauge transformations, generated by
\eqn\unfixexgt{
\ph_1 \r e^{{2 \pi i/ n}} \ph_1,
}
consequently, the low energy physics of our GLSM in the UV limit
is the supersymmetric sigma model on ${\bf C}/{\bf Z}_n$.

\subsec{Deviations from the UV Fixed Point}

Next we consider the deformation of the UV sigma model as $r$ decreases
away from infinity, but is still very large $r \gg 1$. As the GLSM has
${\cal N}=2$ worldsheet supersymmetry for all values of $r$, 
the corresponding
deviation in the sigma model must correspond to adding a linear combination 
of the chiral operators to the worldsheet Lagrangian. The correct deformation
turns out to be 
\eqn\twd{\int d^2z\ e^{-{r + i \theta \over n}} V_{T},}
where $V_{T}$ is the lowest dimensional tachyon vertex
operator in the first twisted sector. In order to see this identification 
more clearly, note that, when $r \gg 1$, \act\ has fractional instantons
(vortices in which the vev of $\ph_n$ winds at infinity \WittenYC). 
Any Euclidean 
correlation function in the theory \act\ may be represented as a sum over 
sectors containing arbitrary numbers of instantons and anti instantons; 
each instanton must also be integrated over the world sheet. On flowing to
low energies, these fractional instantons become twist fields (recall 
that the phase of  $\ph_1$ changes by ${2 \pi \over n}$ on circling a
fractional instanton). The sum over all numbers and positions of
instantons and anti-instantons is thus achieved by perturbing the low
energy sigma model by \twd, where the coefficient of $V_T$ is the action of
a single instanton. 

\subsec{Classical RG flow to the IR}

In this subsection we will compute the sigma model metric corresponding to 
${\bf C/Z}_n$ RG flow, in the classical approximation 
described in subsection
3.2 above.

First we assume that $r$ is positive and solve the D-term condition \ze\
as follows
\eqn\phii
{\ph_{1}=\rho\ e^{i\theta_{1}},\ \ \ \ph_{-n}=
\sq{\f{\rho^2+r}{n}}e^{i\theta_n},}
where the $U(1)$ gauge transformation \ft\ acts as $(\theta_n,\theta_1)
\sim (\theta_n-n\ap,\theta_1+\ap)$.
Then the kinetic terms of the scalar field action become
\eqn\kine{
L=(\partial_{\mu}\rho)^2+\rho^2(\de_{\mu}\theta_1-v_\mu)^2
+\f{\rho^2}{n(\rho^2+r)}(\de_\mu \rho)^2
+\f{\rho^2+t}{n}(\de_{\mu} \theta_n+nv_{\mu})^2.}
The equation of motion of the gauge field in the sigma model limit
$e\to \infty$
leads to
\eqn\vectorf{
v_\mu=\f{\rho^2\de_\mu\theta_1-(\rho^2+r)\de_\mu\theta_n}
{(n+1)\rho^2+nr}.}
After we integrate out the vector field, we get
\eqn\sigmaone{
L=\Bigl(1+\f{\rho^2}{n(\rho^2+r)}\Bigr)(\de_{\mu}\rho)^2
+\f{\rho^2(\rho^2+r)}{n((n+1)\rho^2+nr)}(\de_\mu\theta)^2,}
where we have defined the gauge invariant field
$\theta=n\theta_1+\theta_n$ ($\theta\sim \theta+2\pi$).
In the UV limit $r\to \infty$ (or $r>>\rho$)
this sigma model corresponds to the metric of ${\bf C}/{\bf Z}_n$
\eqn\uvm{
ds^2=d\rho^2+\f{\rho^2}{n^2}d\theta^2.}
In the region $\rho>>r$ we obtain the metric of ${\bf C}/{\bf Z}_{n+1}$
\eqn\metp{
ds^2=\f{n+1}{n} d\rho^2+\f{\rho^2}{n(n+1)}d\theta^2.} As we argue in section
4.2, the appearance of ${\bf C}/{\bf Z}_{n+1}$ is due to our classical 
approximation and the quantum correction should modify 
it such that asymptotic
geometry is given by ${\bf C}/{\bf Z}_n$.

Next we consider the case of negative $r$. The solution to the D-term
condition is given by
\eqn\dtm{
\ph_{1}=\sq{n\rho'^2+|r|} e^{i\theta_1},\ \ \
\ph_{-n}=\rho' e^{i\theta_n}.}
By using the same method as before we obtain the gauge field
\eqn\vectorm{
v_\mu
=\f{(|r|+n\rho'^2)\de_\mu\theta_1-n\rho'^2\de_\mu\theta_n}
{n(n+1)\rho'^2+|r|},}
and the sigma model metric
\eqn\sigmatwo{
ds^2=(1+\f{n^2\rho'^2}{|r|+n\rho'^2})d\rho'^2
+\f{\rho'^2(n\rho'^2+|r|)}{n(n+1)\rho'^2+|r|}d\theta^2.}
In the IR limit $r\to -\infty$ (or $|r|>>\rho'$) 
we obtain the metric of ${\bf C}$
\eqn\irm{
ds^2=d\rho'^2+\rho'^2d\theta^2.}
On the other hand, in the far region $\rho'>>|r|$, the metric looks like
${\bf C}/{\bf Z}_{n+1}$
\eqn\flat{
ds^2=(n+1)d\rho'^2+\f{\rho'^2}{n+1}d\theta^2.}

\subsec{Qualitative Properties of the Exact Flow}

The evolution of the sigma model metric behind occurs in two stages. When  
$r>0$, the metric remains almost constant. More precisely ${\bf C/Z}_n$ 
is deformed to ${\bf C / Z}_{n+1}$ at very large distances; in \sigmaone\
this deformation propagates inwards. This propagation is clearly reversed
by quantum corrections (a bubble of ${{\bf C/Z}_n}$ immersed in 
${\bf C / Z}_{n+1}$ grows rather than shrinks in worldsheet RG flow 
\AdPoSi, \GuHeMiSc.), and is an artifact of our approximation; in the
quantum corrected RG flow we expect the metric to stay constant for $r>0$.

As $r$ turns negative, however, \sigmatwo\ describes a bubble of flat
space which is nucleated at $\rho=0$. This bubble then expands out
in a diffusive manner; its radius grows like 
$R \sim \sqrt{r} \sim  \sqrt{ (n-1) \ln \Lambda}$, in rough agreement with 
the one loop sigma model growth discussed in \AdPoSi,
\GuHeMiSc. Consequently, the evolution of the sigma model described behind 
(together with the minor expected quantum corrections) is summarized in 
Fig. 6, and is in perfect agreement with the APS picture reviewed in B.1. 

\fig{Change of geometry under closed string tachyon in
${\bf C}/{\bf Z}_{n}.$}{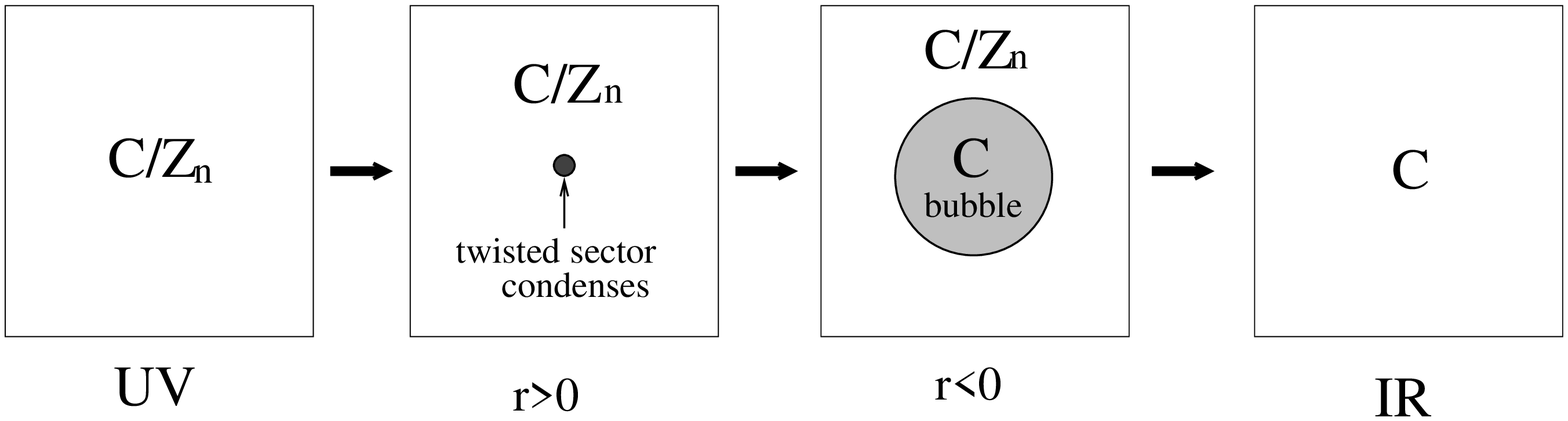}{6truein}

\appendix{C}{Classical Metric in Twisted Circle GLSM}

As in Appendix B, let us compute the classical metric of
the gauged linear sigma model in section 4.3, which has
the kinetic terms
\eqn\kinem{\eqalign{
&L_{kin}=\f{k}{2}(\de_{\mu}{P_1})^2+\f{k}{2}(\de_{\mu}{P_2}+v_\mu)^2
+(\de_{\mu}\rho)^2+\rho^2(\de_{\mu}\theta_1+v_{\mu})^2 \cr
&\ \ \ \ +(\de_{\mu}\rho')^2+\rho'^2(\de_{\mu}\theta_n-nv_{\mu})^2, \cr}}
and the D-term condition $n\rho'^2-\rho^2=kP_1$ (or \zet).
Just in the same way as in B.7 we can integrate out the 
gauge field. After we impose the gauge fixing condition
$\theta_n=0$ from the beginning (just for simplicity)
\eqn\vmum{
v_\mu=-\f{\rho^2 \de_{\mu} \theta_1+\f{k}{2}\de_{\mu}P_2}
{\rho^2+n^2\rho'^2+\f{k}{2}},}
we get the sigma model metric
\eqn\sinem{\eqalign{
&ds^2=(d\rho)^2+(d\rho')^2+\f{k}{2}(dP_1)^2 \cr
&\ \ +\!\f{\rho^2(n^2\rho'^2\!+\!\f{k}{2})(d\theta_1)^2\!-\! k\rho^2
d\theta_1 dP_2\!+\!\f{k}{2}(\rho^2\!+\! n^2\rho'^2)(dP_2)^2}
{\rho^2+n^2\rho'^2+\f{k}{2}}. \cr }}
First we assume $P_1>0$ then we have ($\rho\equiv |\ph_1|$)
\eqn\mets{\eqalign{
ds^2&=(d\rho)^2+\f{k}{2}(dP_1)^2+\f{(\rho d\rho
+\f{k}{2}dP_1)^2}{n(\rho^2+kP_1)}\cr
&+\!\f{(n\rho^4+(nkP_1\!+\!\f{k}{2})\rho^2)(d\theta_1)^2
\!-\!k\rho^2(d\theta_1)(dP_2)\!+\!\f{k}{2}((n+1)\rho^2\!+\!nkP_1)(dP_2)^2}
{(n+1)\rho^2+nkP_1+\f{k}{2}}.\cr}}
In the UV limit $\rho<<k P_1$ we obtain
\eqn\metss{
ds^2=(d\rho)^2+\rho^2(d\theta_1)^2+
\f{k}{2}(dP_1)^2+\f{k}{2}(dP_2)^2,}
which is equivalent to the twisted circle ${\bf C\times S^1}/{\bf Z}_n$
after we take the remained ${\bf Z}_n$ gauge equivalence
$(\theta_1,P_2)\sim (\theta_1+\f{2\pi}{n},P_2+\f{2\pi}{n})$
into account. On the other hand in the opposite limit $\rho^2>>k P_1$
we get
\eqn\mettt{
ds^2\simeq \f{n+1}{n}(d\rho)^2+\f{n}{n+1}\rho^2(d\theta_1)^2
+\f{k}{2}(dP_1)^2+\f{k}{2}(dP_2)^2-\f{k}{n+1}d\theta_1 dP_2,}
which may be described by a twisted circle
 ${(\bf C\times S^1)}/{\bf Z}_{n+1}$. This geometry
is also argued to be an artifact of the classical approximation. If we
take the quantum correction into account, the asymptotic 
region should always 
be the previous twisted circle 
${(\bf C\times S^1)}/{\bf Z}_{n}$.

Finally in the case of $P_1<0$ the result is given by 
($\rho'\equiv |\ph_{-n}|$)
\eqn\metf{\eqalign{
&ds^2=(d\rho')^2+\f{k}{2}(dP_1)^2+\f{(n\rho' d\rho'
+\f{k}{2}dP_1)^2}{n\rho'^2+kP_1}\cr
&+\!\f{(n^2\rho'^2\!+\!\f{k}{2})(n\rho'^2\!+\!kP_1)(d\theta_1)^2
\!-\!k(n\rho'^2\!+\!kP_1)(d\theta_1)(dP_2)\!
+\!\f{k}{2}(n(n+1)\rho'^2\!+\!kP_1)(dP_2)^2}
{n(n+1)\rho'^2+kP_1+\f{k}{2}}.\cr}}
If we take the limit $\rho'^2<<k |P_1|$, then the metric becomes
\eqn\metpo{
ds^2=d\rho'^2+n^2\rho'^2 d\theta_1^2+\f{k}{2}(dP_2-d\theta_1)^2,}
and thus we get the flat space
${\bf C\times S^1}$. This gives the bubble of flat space. It is also easy to
see that
the opposite limit $\rho'^2>>k |P_1|$ corresponds to
${(\bf C\times S^1)}/{\bf Z}_{n+1}$ as before.

\appendix{D}{D-brane Spectrum in dual LG Theory}

\subsec{LG Theory dual to GLSM}

In \HoVa\ Hori and Vafa 
have argued that a gauge linear sigma model with $p$ different
matter chiral multiplets is exactly dual to a Landau Ginzburg model
of $p+1$ interacting twisted chiral multiplets. The first of these
twisted chiral multiplets is simply the gauge multiplet $\Sigma$.
The remaining twisted chiral multiplets are obtained as follows. Each of
the chiral multiplets of the GLSM may be written as $\ph_i=e^{\psi_i}$,
where the imaginary part of $\psi_i$ is periodic with period $2 \pi$.
We now, roughly, T-dualize this GLSM with respect to each of these periodic
variables. The dual angular coordinate pairs up with the real part of $\psi$
and the original fermions to form a twisted chiral multiplet $Y_i$\foot{This
follows from the fact that T-duality reverses the sign of the left moving
part of the angular variable; this exchanges the left moving part of
$\psi_i$ with its complex conjugate, but has no such effect on the right
moving part.}. While the imaginary part of $Y_i$ is a gauge invariant and
supersymmetrized version of the T-dual of the
imaginary part of $\psi $, it turns out that the real part of the twisted
chiral multiplet 
is given by $Y_i+{\bar Y_i}  =2 {\bar \ph_i}e^{2 Q V} \ph_i$
(see eqns 3.18-3.19, \HoVa, for more details). The twisted superpotential
for these fields that follows from the usual T-duality arguments is
$\Sigma (\sum_{i} Q_i Y_i -t)$ where $Q_i$ are the charges of the fields
$\ph_i$. In addition to this classical contribution, the twisted chiral
potential also receives instanton (vortex) contributions; as shown in 
\HoVa\ the exact quantum corrected twisted superpotential is
the classical piece plus $\sum_{i}e^{Y_i}$.

The kinetic terms for these twisted chiral 
multiplets $Y_i$ and $\sigma$ cannot
be computed exactly. However, classically, at low energies ($e \to \infty$)
integrating the $\sigma$ field out simply results in the constraint
$\sum_{i} Q_i Y_i +t=0$ on the $Y_i$ fields. This constraint, the mirror
of the GLSM $D$ term equation, is expected to be exact at low energies.
Consequently, the low energies, we have the Landau Ginzburg theory of
a $p-1$ twisted chiral multiplets ($Y_i$ subject to the constraint described
above) and the twisted chiral superpotential $\sum_{i}e^{Y_i}$.

Now let us apply this procedure to the GLSM \act\ for
${\bf C}/{\bf Z}_n$ in appendix B and
its Liouville version (for twisted circle) in section 3.1. 
Performing the mirror transformation
 we obtain the following LG potential for the former model \Va\
\eqn\sp{
W=e^{-Y_1}+e^{-Y_2}=e^{-nU}+e^{-t/n-U},}
where $Y_1$ and $Y_2$ are dual to $\ph_{1}$ and $\ph_{-n}$ satisfying
the relation $-Y_1+nY_2=t$. In other words, we can write
$Y_1(=nU)=|\ph_{1}|^2+i\phi_{1},\ \ \
 Y_2=|\ph_{-n}|^2+i\phi_{-n},$
where the phases $\phi_{1}$ and 
$\phi_{-n}$ are dual to $\theta_1$ and
$\theta_{n}$.
The periodicity of these fields 
$Y_{1,2}\sim Y_{1,2}+2\pi i$ leads\foot{Here
we used the constraint $-Y_1+nY_2=t$ and thus the periodicity of $U$ is
not $2\pi i/n$ but $2\pi i$.} to the identification
$U\sim U+2\pi i$.
Note that we can see the symmetry ${\rm Im}t(=\theta)\to{\rm Im}t+2\pi i$ 
by considering
the accompanied field redefinition $U\to U-2\pi i /n$.

Now let us consider the change of geometry 
from the viewpoint of the LG theory.
Before the tachyon condensation $(r\to \infty)$, we obtain the simple
potential
\eqn\LGUV{
W=e^{-nU},}
and this has the ${\bf Z}_n$ quantum symmetry 
\eqn\qsLG{
h:U \to U+2\pi i/n,}
which corresponds to the previous quantum symmetry \qs\ in GLSM.
Indeed this potential is the same as the large radius limit ($k\to\infty$) 
of the model 
\HoKa\ and can be
identified with the orbifold ${\bf C}/{\bf Z}_n$.
The closed string tachyon condensation is represented by the second
term of \sp\
and
breaks this symmetry. This term represents
the fractional
instanton (vortex) effect in GLSM and is consistent with our arguments 
around \twd . The
final state after tachyon condensation can be regarded as the flat space
${\bf C}$ as the potential $W\sim e^{-U}$ shows in the $t\to -\infty$ limit.

It is also possible to perform the mirror transformation for the model
discussed in section 3.1
by using the method in \HoKa . Then we obtain the LG potential \DaGuHeMi\
\eqn\LGM{
W=e^{-nU}+e^{-Y_P /n-U},}
where
the fields $Y_P$ denote the dual twisted chiral
field of $P$. One may ask 
how we can see the (dual of)
twisted identification of ${\bf C}/{\bf Z}_{n}\times {\bf S}^1$ 
for the large
$P_1$.
This is understood as follows. Because we have the 
periodicity $2\pi$ for each of
the imaginary parts of $Y_P,Y_1$ and $Y_2$ with the 
constraint $-Y_1+nY_2=Y_P$, we
find not only
the periodicity $U\sim U+2\pi i{\bf Z}$ and $Y_p\sim Y_p+2n\pi i{\bf Z}$
but also the ${\bf Z}_n$ identification
\eqn\twi{
(U,Y_P)\sim (U+2\pi i/n,Y_P-2\pi i).}
This projection \twi\ just corresponds to the twisted identification (see
Appendix E).
Because of this monodromy the true radius is 
$\ti{R}_{UV}=\f{n\sqrt{\al}}{\sqrt{k}}$.
Thus it is obvious that we obtain the twisted circle in the UV limit.
On the other hand, in the IR limit the dominant term becomes the second
term and thus we get the flat space\foot{
If we consider $P_1$ takes a large negative value,
then we should use the good
variables $Y_1,Y_p$ as $W=e^{-nY_1+Y_p}+e^{-Y_1}$ and thus
we have no such twisted
identification ($\ti{R}_{IR}=\sqrt{\f{\al}{k}}$).}.

\subsec{Description of D-branes in LG Theory dual to Orbifold}

In this subsection we will investigate the D-brane spectrum 
of ${\bf C/ Z}_n$ in the dual Landau Ginzburg model \LGUV. 
We will use the constraints from ${\cal N}=2$ supersymmetry 
 \HoIqVa \Goa\ in what follows.

We concentrate on D1-branes which preserve the A-type supersymmetry 
(A-brane). The general arguments of \HoIqVa \Goa, demonstrate that 
such D-branes have a (middle dimensional) Lagrangian sub manifold as 
their world-volume; further the imaginary part of the superpotential is 
constant along this world-volume. The first condition is satisfied by 
any curve in the $U$ plane. The second one is crucial and is less easily
satisfied. We first consider the 1-branes that are mirror (T-dual) to the 
GLSM D2-branes $|m\lb_{D2}$ of subsection 4.1. Note that
we T-dualize along the Im$U$ direction to map between these pictures, 
consequently the branes we are after are localized in the $Im U$
direction. They are given by \foot{Even though
one can consider other possibilities of curved world-line, they do not
satisfy the boundary conformal invariance in the UV limit.}
\eqn\lgdd{
{\rm{Im}}U=\f{2\pi im}{n},}
where the numbers $m=0,1,\ddd,n-1$ label the types of 
D1-branes\foot{One may note that the supersymmetric
condition allows the half-integer value of $m$. There are related
to integer $m$ by the operation $(-1)^{F_L}:U\to -U$. Since the
boundary state is invariant under GSO projection $\f{1+(-1)^{F_L}}{2}$,
we should add the brane $m+n/2$ to $m$ in \lgdd.
This complication does not change our arguments later and we will not 
explicitly mention it below.}. We denote this
world-volume \lgdd\ by $\gamma_{m}$ (see Fig.7 below).
Indeed the ${\bf Z}_{n}$ quantum symmetry action
$h:{\rm Im}U \to {\rm Im}U+2\pi i/n$ 
generates the other D1-branes in the same way as in GLSM.

\fig{The LG description of D2-brane.}{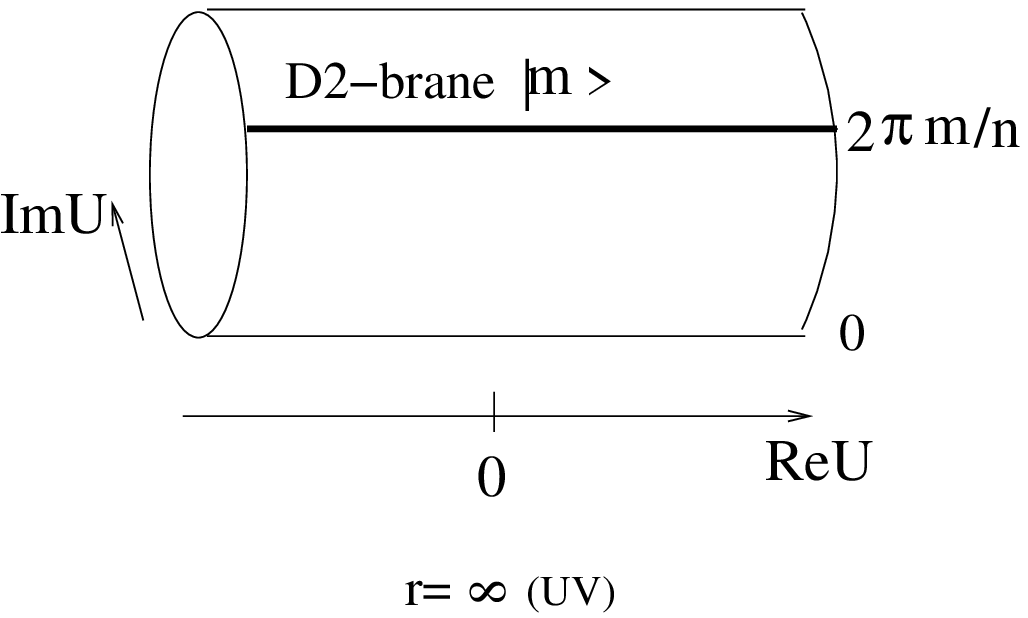}{2.5truein}

We now turn to a discussion of D1-branes that are dual to D0-branes. 
The direct determination of such branes is complicated by the following
fact. The Landau Ginzburg model that reduces to 
${{\bf C/Z}_n}$ has vanishing kinetic term (this is the $k\to \infty$ limit
 in \HoKa). The Liouville potential, is, consequently effectively 
infinite when Re$U$ is negative, and is effectively zero 
when Re$U$ is positive. For any finite value of the the LG kinetic term 
(if $k$ in \HoKa\ is not strictly infinite), the wrapped 1-branes of the 
Landau Ginzburg model \sp\ are attracted to, and stuck at, the 
strong coupling region. Strictly in the $k \to \infty$ limit this 1-brane 
is free to wander in the bulk. Thus we do not expect to find any ${\bf
C/Z_n}$ brane, localized in the $ReU$ direction, using the usual 
Landau Ginzburg techniques (which work when the LG kinetic term is 
nonsingular). Nonetheless it is not difficult to build a 
qualitative picture of
the states we expect. 

As discussed in section 5, GLSM 0-branes may be obtained from 
GLSM $D2-\overline{D2}$ by the open string tachyon condensation. 
A bulk D0-brane is a vortex on a D2 and an anti D2-brane of the same sort; 
consequently it is natural to conjecture that the world volume of such a
brane is a circle along the Im$U$ direction (see the left-hand side of
Fig.8). Indeed it has the correct
D0-brane charge after the T-duality and is obviously movable away from the
fixed point in the $k \to \infty$ limit\foot{
One may think that this configuration of a D0-brane breaks the A-type
supersymmetry because it does not satisfies the condition of
superpotential mentioned before. However, our non-compact orbifold
corresponds to the limit of vanishing the kinetic term of the
Liouville theory as we have noted. Thus the Liouville
potential can be neglected for Re$U>0$ and it will keep the A-type
supersymmetry. This is also equivalent to the fact that the dilaton
is constant in the limit $k\to\infty$.}.
On the other hand, we can generate a fractional D0-brane
$|m\lb_{D0}$ from a D2 $|m+1 \lb_{D2}$ and a $\overline{D2}$
$|\bar{m} \lb_{\bar{D2}}$. In this case we argue that its world-volume
after the open string tachyon condensation is (approximately)
given by (see the right-hand side of Fig.8)
\eqn\lgdez{
\{U=x\ |x<0\}\cup \{U=\f{2\pi im}{n}+i\theta\ |0\leq
\theta\leq \f{2\pi}{n}\}\cup
\{U=y+\f{2\pi i}{n}\ |y<0\}.}
It seems reasonable that annihilation (open string tachyon condensation)
of the $|\bar{m} \lb_{\bar{D2}}$ occurs only for
positive Re$U$ because the existence of the Liouville potential; thus this
open string tachyon condensation leaves behind a remnant stuck at $U=0$; 
this remnant is the T-dual of the fractional 0-brane. Note, of course,
that it winds a fraction of the Im$U$ circle. 

\fig{The LG description of D0-brane.}{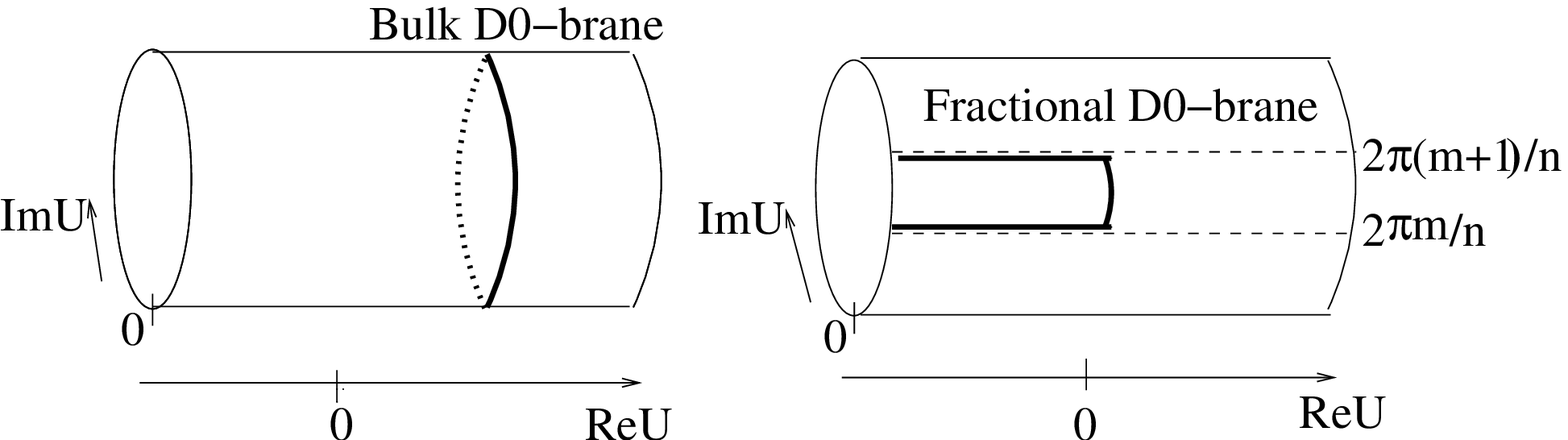}{4truein}

\subsec{Computation of RR-charges}

To check these interpretations, let us compute the coupling to
the RR-sector ground state by using the information of superpotential.

In the papers \HoIqVa \Goa\  D1-branes in
the LG theory
of $W=\Phi^{k+2}$, which is equivalent to the ${\cal N}=2$ minimal 
model in the
IR limit, were
analyzed and
we can use this result by performing the change of the variable
$\Phi\to e^{-U}$.

First we start with the D1-brane given by \lgdd . The couplings to
the $k^{th}$ twisted RR-sector ground state 
(${\cal N}=2$ chiral state) can be expressed \foot{
Here the sector $k=0$ corresponds to the untwisted sector and its
coefficient is divergent due to the infinite volume. Below we can
omit this sector because it does not contribute to
the present calculations.} by the following `boundary state'
\eqn\bdd{
|m \lb_{D2}=\sum_{k=0}^{n-1}\f{c^{0}_{k,m}}{N_k}|e^{-kU}\lb,}
where we can compute the coefficient $c^{0}_{k,m}$ as follows
\eqna\cofb
$$\eqalignno{
c^{0}_{k,m}
&=\int_{\gamma_{m}}dUe^{-kU}e^{-W} \cr
&=e^{-2\pi im k/n}\int^{\infty}_{-\infty} dx e^{-kx}e^{-e^{-nx}}=
\f{e^{-2\pi im k/n}}{n}\ \Gamma(k/n).\cr}$$
The normalization $N_k$ of the RR-sector ground state $|e^{-kU}\lb$
can also
determined by using the intersection number (see the later arguments and
also \HoIqVa (p.50) for more detail)
\eqn\ints{
N_k=\la \ov{e^{-kU}}|e^{-kU}\lb=\f{2\sin (\pi k/n)}{n}
\Gamma(k/n)^2.}
Thus we get the normalized couplings
\eqn\lgdtb{
c_{k,m}(=\f{c^{0}_{k,m}}{\sq{N_k}})=
\f{1}{\sq{2n\sin(\pi k/n)}}e^{-2\pi im k/n}.}
This result matches 
exactly with the D2-brane
boundary state in
the orbifold theory \ta\ including the normalization.

Next we turn to D1-branes which are dual to D0-branes in GLSM.
Their couplings to the RR-sector ground states $c'_{k,\ap}$ do not change
under open string tachyon condensation 
and thus we can compute it as follows
\eqn\lgdzb{
c'_{k,m}=c_{k,m+1}-c_{k,m}=
\sq{\f{2\sin(\pi k/n)}{n}}e^{-\pi i(k/n+1/2)}\cdot e^{-2\pi im k/n}.}
Thus we again find the exact agreement with the result of the
boundary states in
the orbifold theory \DiGo\ \Bi\ up to an irrelevant phase factor.
Note also that the intersection numbers $I_{m,m'}$
(or equally open string
Witten index Tr$_{R}(-1)^F$ \DoFi) of these D1-branes are
given by the overlap of the two boundary state with the
insertion\foot{
In the case of ${\cal N}=2$ minimal model we should take
$(-1)^{F_L}=ie^{\pi ik/n}$ and the index is slightly different from our
model.} of
$(-1)^{F_L}=ie^{\pi ik}$ as follows
\eqna\indo
$$\eqalignno{
\la m|(-1)^{F_L}|m' \lb_{D0}
&=\sum_{k=1}^{n-1}ie^{\pi ik/n}\cdot \f{2\sin(\pi k/n)}{n}
\cdot e^{\f{2\pi i}{n}(m-m')} \cr
&=\delta_{m,m'+l}-\delta_{m,m'+l+1},\cr}$$
and indeed reproduce the quiver theory 
result\foot{This is easy to see from the fact that the ${\bf Z}_n$
projection is given by $e^{\f{2\pi i}{n}(m-m')}\cdot 
e^{\f{2\pi i(n+1)}{n}s_1}=1$,
where $s_1=\pm \f{1}{2}$ is the (spacetime) spin of fermionic
ground states in ${\bf C}/{\bf Z}_{n}$.}.

\subsec{Speculations on D-branes in the LG Model}

Consider the LG dual to the ${\bf C/Z}_n$ 2-brane $|0\rangle$ described in 
subsection 4.2. As we have described above, the Landau Ginzburg description
of this state is a 1-brane located at Im$U$=0 (assuming $\theta=0$). 
In fact Im$U$=0 continues to 
preserve ${\cal N}=2$ supersymmetry even upon turning on the tachyon
superpotential in \sp; so the mirror to $|0\rangle$ evolves into the mirror
of a bulk 2-brane in flat space, in agreement with our previous results.

The situation is not as simple for other 2-branes in \lgdd. On turning on
the tachyon contribution to \sp, the world line of these branes begins to
bend inwards towards Im$U$=0 (see Fig. 9) (this follows from the condition
that the imaginary part of the superpotential remain constant). However, as
the closed string tachyon is localized, at any finite $r$ the effect of
tachyon condensation should be restricted to the region 
$Re$U$ \approx \sqrt{|r|}$. 
Consequently the world line of the 1-brane may be
expected to curve back up to its initial value (see the right end of
Fig. 9). This configuration naively breaks ${\cal N}=2$ 
worldsheet supersymmetry,
but, as in subsection D.2, this breaking may vanish in the $k \to \infty$
limit. If these speculations are indeed true, they would also explain
the disappearance of the fractional 0-branes \lgdez\ under tachyon 
condensation. However we leave the verification (or otherwise) of these
speculations to future work.

\fig{The D2-brane in LG description after the RG-flow.
	We show the world-volume of both D2-branes $|m \lb_{D2}$ and
$|0 \lb_{D2}$. Even though the former D2-brane will be deformed
for $-|r|/n(n-1)<$Re$U<|r|/n$, where the Liouville potential
$e^{-U}$ dominates the superpotential, it does not change
       in the asymptotic region Re$U>|r|/n$. Finally we can see that
in the limit $r\to -\infty$ all these fractional
	D2-branes will become the same usual D2-brane in flat space.
       (In this picture we assumed $\theta=0$ for simplicity.)}
{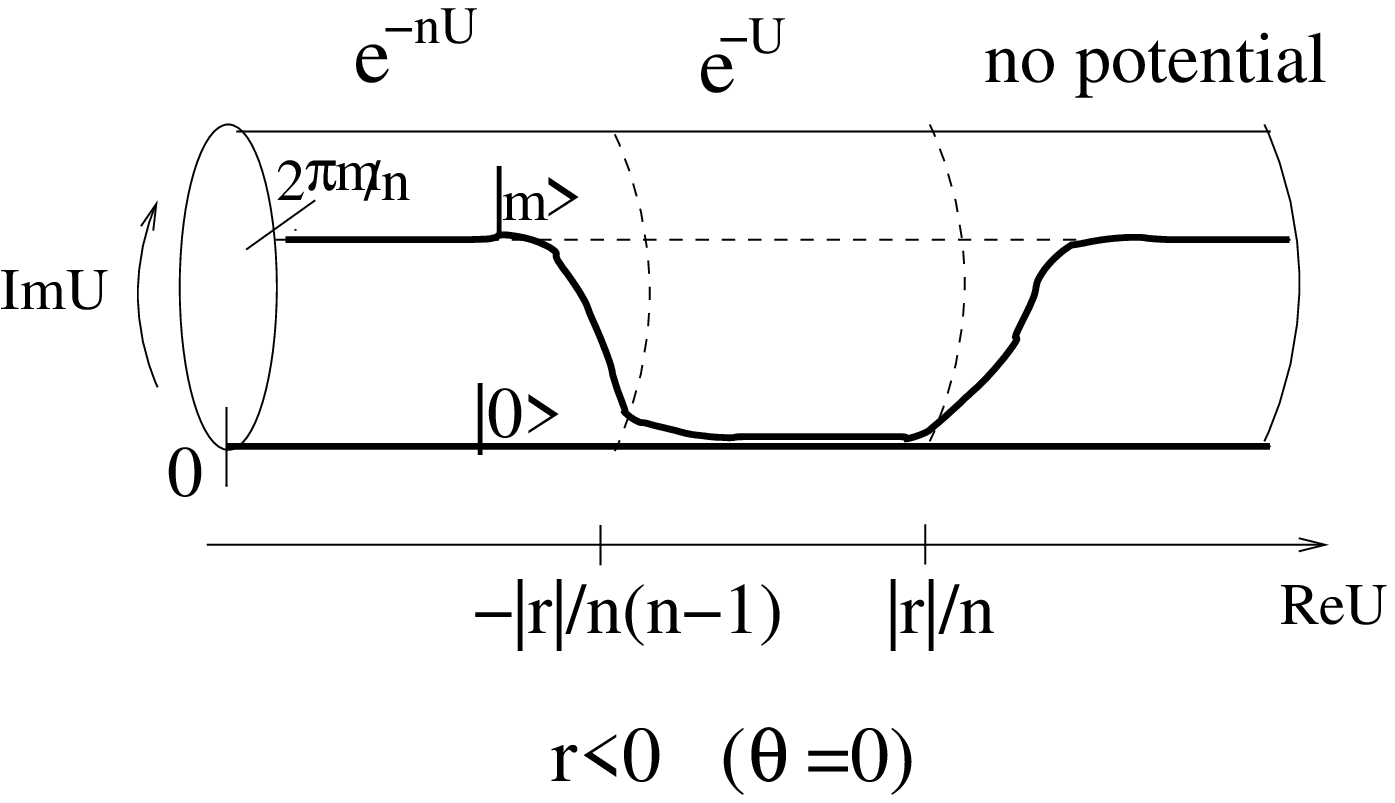}{3truein}

\appendix{E}{Useful T-duality Relations}

Here we would like to summarize the various T-dual expressions
of the twisted circle $({\bf C}\times {\bf S^1})/{\bf Z}_n$ model.
We denote the ${\bf Z}_n$ projection
by $g=\exp(\f{2\pi i(n+1)}{n}J_{12})$ and its dual ${\bf Z}_n$
quantum symmetry \foot{Note that this includes a
particular example of $g=(-1)^{F_S}$ and $h=(-1)^{F_L}$ for ${\cal N}=2$
(see also \bega, \CoGu, \AdPoSi ).}
 by $h$ \qs . The $1/n$ shift is also
represented by $\sigma_{1/n}$ and its dual operator by
$\tilde{\sigma}_{1/n}$. Then we obtain the following equivalent
descriptions
\eqna\tdualr
$$\eqalignno{
&{\rm{IIA(B)\ \ \  Melvin\ \  Model\ \  (radius}}\ \  R\ \  {\rm{and}} \ \
\zeta(=qR)=(n+1)/n) &\tdualr a\cr
& \ \ \ \simeq
{\rm{IIA(B)\ \ on\ \  }} [{\bf C}\times {\bf S^1}\ \
({\rm{radius}}\ \  nR)]/(g\cdot \sigma_{1/n})\ \ \
{\rm{(=twisted\ \  circle\ \ )}}  &\tdualr  b\cr
& \ \ \ \simeq {\rm{IIA(B)\ \  on\ \  }} [{\bf C}/{\bf Z}_n
\times {\bf S^1}\ \ ({\rm{radius }}\ \ R)]/
(h\cdot\ti{\sigma}_{1/n})  &\tdualr c\cr
& \ \ \ \simeq
{\rm{IIB(A)\ \  Melvin\  Model\ \  (radius\ \  1/R\ \
 and}}\ \  \beta\al R=(n+1)/n)  &\tdualr  d\cr
& \ \ \ \simeq {\rm{IIB(A)\ \  on\ \  }} [{\bf C}\times {\bf S^1}\ \
({\rm{radius }}\ \  1/nR)]/(g\cdot \ti{\sigma}_{1/n})  &\tdualr e\cr
& \ \ \ \simeq {\rm{IIB(A)\ \  on\ \  }} [{\bf C}/{\bf Z}_n
\times {\bf S^1}\ \ ({\rm{radius }} \ 1/R)]/
(h\cdot \sigma_{1/n}), &\tdualr f \cr }$$
where the parameter $\beta$ corresponds to the B-field flux T-dual to the 
twisted parameter $q=\zeta/R$ (see \tse \TaUeo).
These are all explained as T-duality relations.

Finally let us consider the relation to LG theory. As we have seen, the UV
region ($Y_p\to\infty$) of the
LG theory \LGM\ is equivalent to the twisted circle. This is
understood in the above picture as follows. This LG theory is defined by
the superpotential \LGUV , which is equivalent to
the orbifold ${\bf C}/{\bf Z}_n$, and by the twisted
identification \twi\
\eqn\wlg{
{\rm{typeIIA(B)\ on}}\   [{\bf C}/{\bf Z}_n
\times {\bf S^1}\ \ ({\rm{radius }}\  1/R)]/
(g'\cdot\sigma_{1/n}),}
where the ${\bf Z}_n$ action $g'$ denotes the shift $U\to U+2\pi i/n$.
If we take T-duality in the angular direction of ${\bf C}$, then
we have the projection $h$ instead of $g'$.
Therefore this model is indeed
equivalent to the expression \tdualr{f}.

\listrefs

\end